**Space-Charge-Driven Nonlinear Charge Transport in Silicon Reconfigurable Nonlinear-Processing Units**

*Jonas Kareem*, *Lorenzo Cassola*, *Reinier J. C. Cool*, *Janiek I. van Slooten*, *Ray J. E. Hueting*, *Peter A. Bobbert*, *Wilfred G. van der Wiel*[*]

J. Kareem, L. Cassola, R. J. C. Cool, J. I. van Slooten, P. A. Bobbert, W. G. van der Wiel
NanoElectronics Group, MESA+ Institute, and Center for Brain-Inspired Computing (BRAINS), University of Twente, P.O. Box 217, 7500 AE Enschede, The Netherlands
E-mail: w.g.vanderwiel@utwente.nl

R. J. C. Cool, W. G. van der Wiel
Institute of Physics, University of Münster, 48149 Münster, Germany

R. J. E. Hueting
Power Electronics Group, MESA+ Institute, University of Twente, P.O. Box 217, 7500 AE Enschede, The Netherlands

P. A. Bobbert
Materials to Optoelectronic Devices and Eindhoven Institute for Renewable Energy Systems, Eindhoven University of Technology, P.O. Box 513, 5600 MB Eindhoven, The Netherlands

Funding: This publication is part of the project NL-ECO: Netherlands Initiative for Energy-Efficient Computing (with project number NWA. 1389.20.140) of the NWA research programme Research Along Routes by Consortia which is financed by the Dutch Research Council (NWO). We acknowledge financial support from the HYBRAIN project funded by the European Union's Horizon Europe research and innovation programme under grant agreement no. 101046878. This work was further funded by the Deutsche Forschungsgemeinschaft (DFG, German Research Foundation) grant no. SFB 1459/2 2025–433682494.

Keywords: transport phenomena, unconventional computing, semiconductor device modeling, space charge, semiconductor device fabrication, silicon, nonlinear system

**Abstract**

Reconfigurable nonlinear-processing units (RNPUs) are multi-terminal electronic devices that act as computational primitives, exploiting intrinsic nonlinear charge transport combined with electrostatic tunability. Silicon-based realizations provide a scalable and technologically relevant platform toward unconventional computing hardware, yet the physical origin of their room-temperature nonlinearity has remained unresolved. Here, we demonstrate room-temperature operation in both boron- and arsenic-doped silicon RNPUs and show, using temperature- and length-dependent measurements supported by TCAD simulations, that charge transport is governed by space-charge effects. Interface trap states strongly suppress the equilibrium carrier density, while the functional nonlinearity arises from the competition between injected carriers and ionized dopants. The resulting transport evolves from Ohmic to strongly nonlinear and space-charge-limited current regimes, as evidenced by voltage and length scaling. The opposite-polarity background doping is shown to control the onset and strength of the nonlinearity, producing behavior beyond the quadratic dependence of the classical Mott–Gurney law. Agreement between experiment and simulation supports that the spatial distribution of injected carriers and fixed charge governs the electric-field profile and device response. These results establish a space-charge-based framework for RNPUs that does not require disorder or hopping transport, and provide design guidelines for scalable, CMOS-compatible nonlinear computing hardware.

## 1. Introduction

In emerging computing hardware paradigms, including neuromorphic and broader unconventional electronics, nonlinear current-voltage response and reconfigurability of conduction pathways are central to their functionality [1]. For example, resistive-switching memristors rely on filamentary or trap-mediated transport to achieve reconfigurable states [2–6], while reconfigurable transistors exploit tunable barrier or junction conduction mechanisms for adaptive logic and memory functions [7–9]. Silicon-based reconfigurable nonlinear-processing units (RNPUs) similarly leverage intrinsic nonlinearity and tunability, enabling the solution of linearly inseparable classification problems [10, 11] and implementation in hardware neural-network emulators [12]. More recently, they have served in temporal signal processing tasks [13] and as computing primitives in physical Kolmogorov-Arnold Networks [14, 15]. Their realization in silicon and their room-temperature operation furthermore provide a direct pathway toward CMOS-compatible unconventional electronics.

Despite this progress, the underlying charge transport mechanism in silicon-based RNPUs has remained insufficiently understood. Early RNPUs operated at cryogenic (liquid-nitrogen) temperature and their behavior was interpreted in terms of hopping-mediated transport in a disordered dopant network [10, 16, 17]. However, an unusually high dopant activation energy needed to be assumed to explain the temperature dependence of the charge transport [13, 18], suggesting that transport mechanisms other than the proposed 2D variable-range hopping (VRH) via dopant states may be relevant. Recently demonstrated RNPUs [13] operating at room temperature further point toward transport mechanisms beyond hopping. While VRH can in principle occur via dopant states even in crystalline silicon [19–21], such mechanisms are typically most relevant at low temperatures, where carriers remain localized. At room temperature, thermally activated transport via extended states increasingly dominates over localized hopping pathways, making a hopping-dominated interpretation less compelling for the present devices.

Here, we elucidate the origin of nonlinear charge transport in crystalline silicon RNPUs based on both boron- and arsenic-doped devices through experimental current–voltage ($I$–$V$) characterization, including channel-length and temperature dependence, complemented by commercial technology computer-aided design (TCAD) simulations [22]. We show that space-charge effects in the presence of interface states and, crucially, background doping are key to charge carrier transport at room temperature. Unless stated otherwise, the presented results focus on boron-doped devices, while analogous behavior is observed in arsenic-based RNPUs, confirming the generality of the proposed space-charge-driven transport framework across doping species. This insight enables the optimization of future device generations, including implementations aligned with standard CMOS processing. Importantly, we demonstrate that tunable nonlinearity and negative differential resistance (NDR) can be realized without invoking disorder, underscoring the potential for reproducible and scalable integration. The largely unexplored use of space-charge effects in silicon for unconventional electronics points to a promising direction for such architectures.

## 2. Results

### 2.1. Room-temperature nonlinearity and functionality

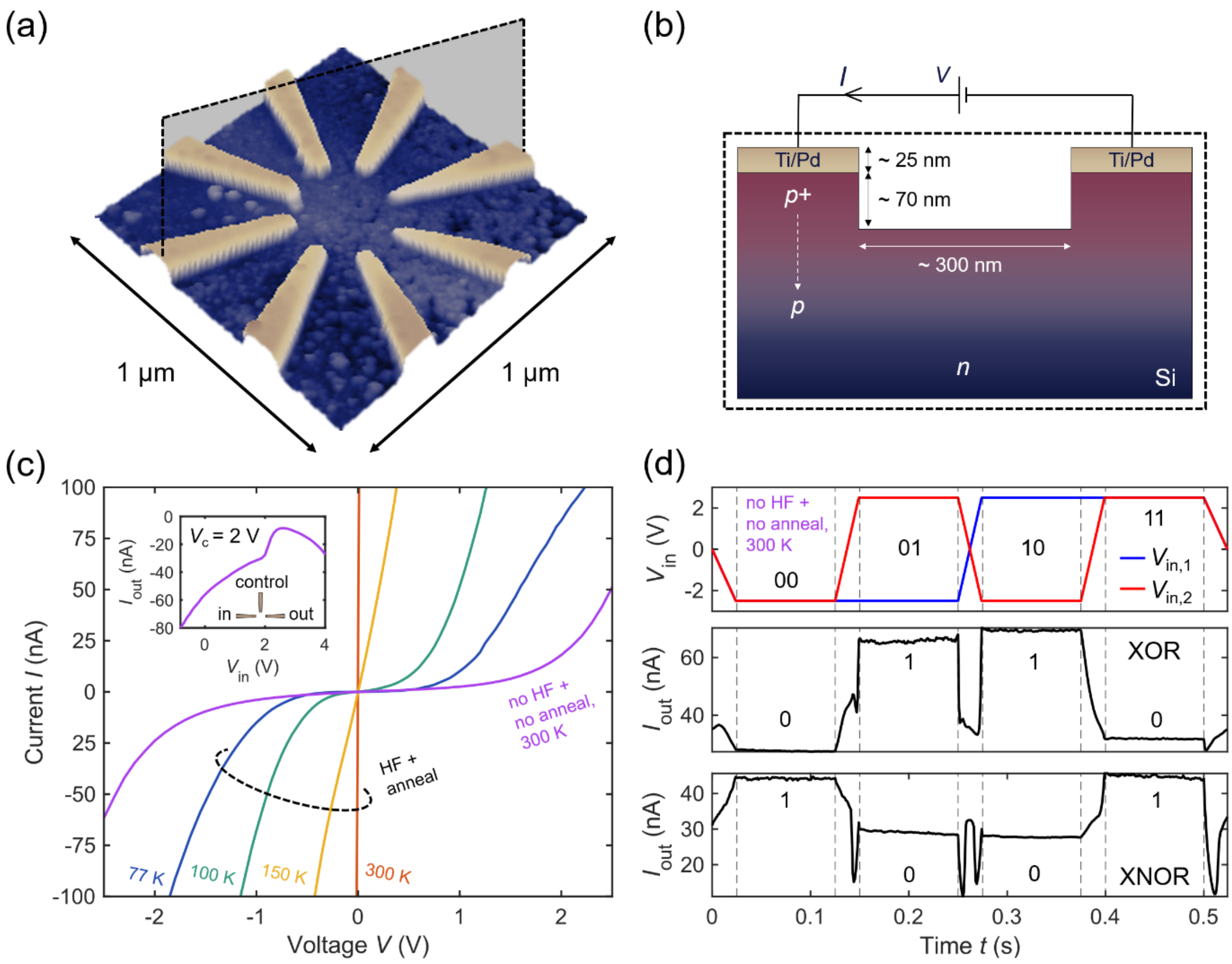

**Figure 1**. RNPU characteristics and logic functionality. (a) Atomic force microscope image of an RNPU. (b) Schematic cross section of a boron-doped RNPU highlighting characteristic device dimensions. (c) Two-terminal *I*–*V* characteristics of boron-doped silicon reconfigurable nonlinear-processing units (RNPUs), comparing untreated devices (no HF cleaning or annealing after dry etching, see text) with previously reported treated devices [10], at various temperatures. Untreated devices exhibit pronounced nonlinear behavior at room temperature, in contrast to the predominantly linear response of treated devices. Inset: An additional current-carrying electrode ($V_c = 2$ V, $V_{out} = 0$ V) enables negative differential resistance (NDR). Additional (otherwise floating) electrodes further

enhance tunability. (d) Untreated device demonstrating X(N)OR Boolean logic. Output current $I_{\text{out}}$ measured in response to voltages $V_{\text{in},1}$ and $V_{\text{in},2}$.

We consider silicon-based, 8-electrode RNPUs, hereafter simply referred to as RNPUs, with either As or B ion implantation doping (**Figure 1**a, **Figure S1**) [10, 13]. We use standard (100) silicon wafers that are lightly doped (~$10^{15}$ $\text{cm}^{-3}$), with a doping type opposite to that of the implanted dopants (see Section 4.1). The ion implantation results in a doping gradient [10], with degenerate doping near the surface and a gradual decrease with increasing distance to the surface (**Figure S2**). Ohmic contacts are defined on the degenerately doped surface via electron-beam lithography and metal evaporation. The devices are then wet-etched in hydrofluoric acid (HF), which removes the native oxide; the silicon surface re-oxidizes upon exposure to ambient conditions. The silicon is subsequently dry-etched to reduce the doping (and hence the free carrier concentration) between the terminals, exploiting the decay of the doping concentration with depth (Figure 1b). We use reactive-ion etching (RIE), which provides an anisotropic etch profile through the deposition of fluorocarbon species on the sidewalls (**Figure S3**). The metallic electrodes thus act as a hard mask, while the anisotropic etch preserves the Ohmic contacts. Thus, devices are fabricated within a larger implantation window (Figure S1), rather than using local contact implantation. More details on the fabrication of devices are provided in Section 4.1.

Previously investigated cryogenic RNPUs operated at 77 K, while room-temperature operation was achieved by applying a positive voltage to the lightly *n*-doped substrate of boron-implanted devices [10]. This was explained by the widening of the depletion region at the *p*-*n* junction between the top layer and the substrate, accompanied by a suppression of band conduction. In the cryogenic RNPUs of Ref. [10], a post-RIE HF etch was first applied to remove fluorocarbon residues resulting from RIE (**Figure S4**). However, this treatment leaves the silicon surface exposed to air, likely increasing the surface defect density [23], as reflected by the significant increase in resistance observed in the investigated devices following the standard HF cleaning step (**Figure S5**). A subsequent low-temperature anneal (180–250°C) was therefore introduced to reduce the surface defect density [23] and partially restore conductance. The resulting increase in conductance leads to strongly Ohmic behavior at room temperature, requiring operation at lower temperatures to suppress the free carrier density.

By omitting the HF cleaning step, the low-temperature anneal is no longer required, eliminating the need for cryogenic operation. This enables room-temperature RNPUs with nonlinear *I–V* characteristics similar to those observed previously at lower temperatures (Figure 1c). By simply biasing an additional control electrode with voltage $V_c$, the output current exhibits a complex dependence on the input voltage, including NDR. This property is essential for RNPU functionality, as it introduces effective negative feedback (inhibitive response) [24–27], enabling the realization of nonlinearly separable operations such as the XOR gate. A more detailed investigation of the NDR behavior and device functionality will be presented elsewhere [28].

Accordingly, we demonstrate basic room-temperature functionality by realizing linearly inseparable X(N)OR gates (Figure 1d). Compared to cryogenic RNPUs [10], we observe improved signal-to-noise ratio (SNR) and output-state discrimination, enabled by the use of slightly higher control electrode voltages (up to 2.5 V) associated with the onset of strong nonlinearity. Other Boolean logic gates are also readily solvable (**Figure S6**).

## 2.2. Role of traps at equilibrium

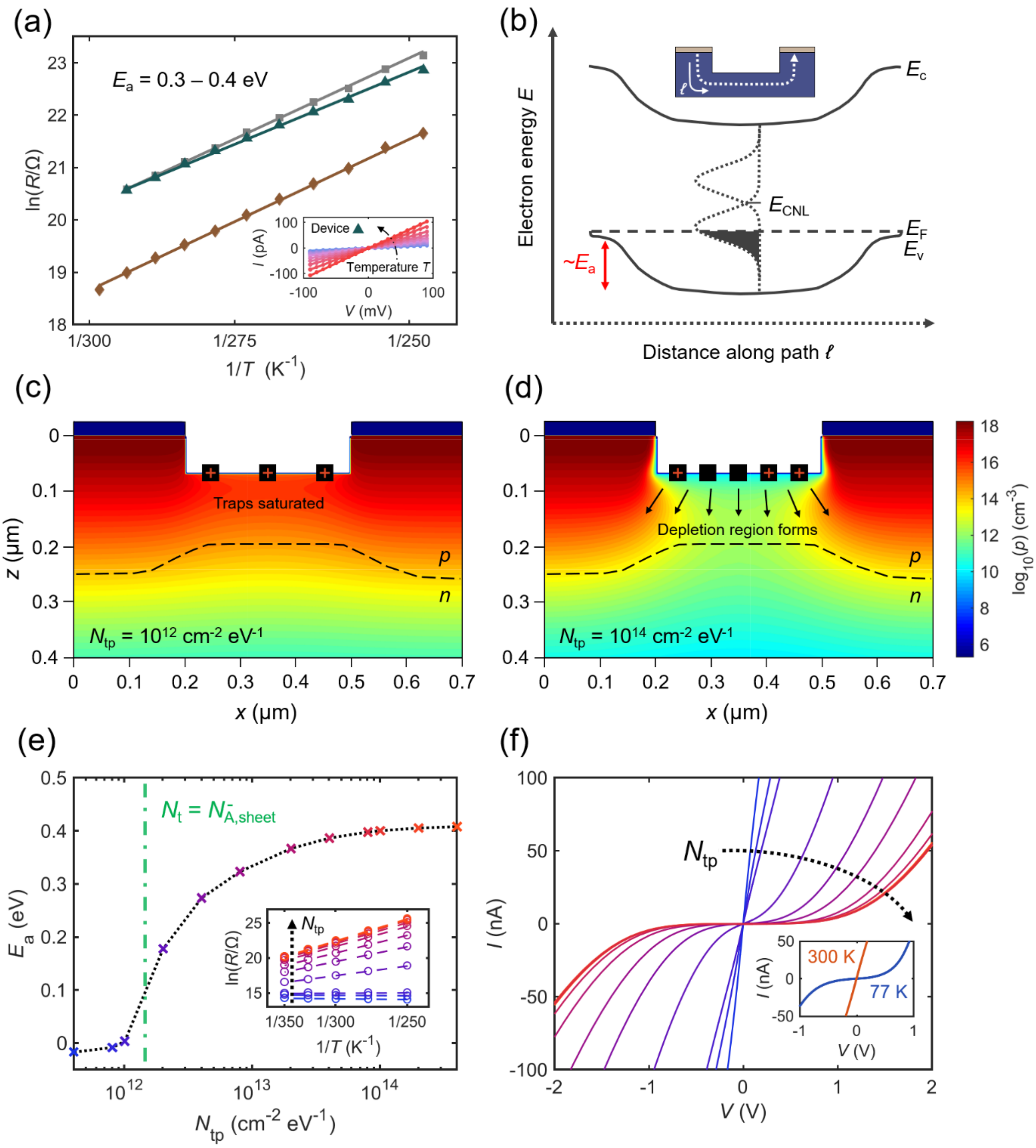

**Figure 2**. Trap-induced reduction of the free carrier concentration. (a) Arrhenius plots of low-bias resistance $R$ (Ω) for boron-implanted silicon RNPUs with different etch times: diamonds (1.5 min), squares (3 min), and triangles (4.5 min), with extracted activation energies $E_a$. Inset: Low-bias $I$–$V$ characteristics of a single RNPU (triangle) versus temperature (opposite contacts). (b) Equilibrium band diagram along current path $\ell$ near the silicon surface (inset) in a boron-implanted device. Surface trap states (see text) pin the Fermi level $E_F$ (indicated by dashed line) near the charge-neutrality level $E_{CNL}$

between the conduction and valence band edges ($E_\mathrm{c}$ and $E_\mathrm{v}$), suppressing free carriers and forming a hole barrier with height $E_\mathrm{a}$. (c-d) Technology computer-aided design (TCAD) contour plots of free hole concentration for $N_\mathrm{tp} = 1 \times 10^{12}\ \mathrm{cm^{-2}\ eV^{-1}}$ and $1 \times 10^{14}\ \mathrm{cm^{-2}\ eV^{-1}}$, respectively. Dashed line indicates location of *p-n* junction. (e) TCAD simulation of activation energy $E_\mathrm{a}$ as a function of peak interface trap density per unit energy $N_\mathrm{tp}$. The dash-dotted line indicates the transition from a very low to a high $E_\mathrm{a}$. Inset: Arrhenius plots of low-bias resistance $R$ for increasing $N_\mathrm{tp}$. (f) Simulated *I–V* characteristics for increasing $N_\mathrm{tp}$. Beyond $N_\mathrm{tp} \sim 1.0 \times 10^{14}\ \mathrm{cm^{-2}\ eV^{-1}}$ the *I–V* characteristics do not change, in agreement with (e). Inset: *I–V* characteristics at 77 K and 300 K for $N_\mathrm{tp} = 1 \times 10^{12}\ \mathrm{cm^{-2}\ eV^{-1}}$. Slight asymmetry in the *I–V* characteristics at 77 K is attributed to numerical error.

In both cryogenic [10, 18] and room-temperature RNPUs (**Figure 2**a), we observe a distinctly high activation energy $E_\mathrm{a}$ for low-bias transport between two opposite contacts, typically in the range of 0.3–0.4 eV for relevant etch depths (see Methods). This suggests that their equilibrium transport mechanism shares a common origin. The resistance $R$ in Figure 2a shows Arrhenius-like behavior, characteristic of a thermally activated process:

$$R = R_0 \exp\left(\frac{E_\mathrm{a}}{k_\mathrm{B}T}\right), \tag{1}$$

with $R_0$ a prefactor collecting all non-activated contributions. We propose that the large $E_\mathrm{a}$ arises from a high density of surface states resulting in Fermi-level pinning close to mid-gap (Figure 2b). The surface damage induced by RIE is expected to significantly increase the density of dangling bonds at the silicon/native oxide interface, which act as traps. The predominant interface defects on (100) Si are $P_\mathrm{b}$ centers, namely $P_\mathrm{b0}$ and $P_\mathrm{b1}$ centers [29, 30], with $P_\mathrm{b0}$ typically dominating. Due to the amphoteric nature of these traps [31], introducing both donor- and acceptor-like states in the bandgap, both *n*-type (arsenic-implanted) and *p*-type (boron-implanted) devices exhibit high activation energies. The Fermi level pins near the charge-neutrality level $E_{CNL}$ [32], which depends sensitively on surface properties [33] and is likely governed by defect formation during and immediately after dry etching. These defects compensate the implanted dopants by trapping the free carriers they introduce, thereby significantly reducing the carrier concentration in the bulk of the device.

We confirm this scenario using Silvaco TCAD device simulations [22] (see Section 4.5), considering a 2D geometry that approximates a two-terminal RNPU. Modelling the $P_b$ centers as a pair of Gaussian trap distributions at the silicon/oxide interface (**Figure S7**) shows that the Fermi level is pinned near mid-gap between the contacts (Figure 2b) and that the (equilibrium) carrier concentration is indeed significantly reduced in the etched region, extending from the surface into the bulk (**Figure S8**). The free hole concentration is approximately six orders of magnitude lower than expected from the implanted dopant concentration (Figure S2). At the silicon surface, both $n$ and $p$ approach the intrinsic carrier concentration $n_i$ because the interface states capture carriers introduced by the dopants, causing the material to behave effectively as intrinsic silicon. Throughout the silicon depth, the mass-action law

$$pn = n_i^2 \tag{2}$$

holds, with the hole concentration peaking approximately 100 nm below the surface before decreasing again due to the implanted dopant profile. The exact carrier distribution depends on the capture cross-sections of the defect centers, as described by Equation (18). Consequently, the hole concentration reaches its minimum at the silicon surface, where carrier trapping is strongest.

In combination with the degenerately doped regions beneath the electrodes, this depletion of carriers creates a potential barrier between the contacts, analogous to a FET [34], with a height corresponding to the measured activation energy. In RNPUs, this barrier is not controlled by a capacitive gate, but by the electric field arising from the applied voltages and injected carriers. The distribution of interface states within the bandgap determines the position at which the Fermi level becomes pinned in the etched region and therefore directly controls the barrier height. Sufficiently high interface trap densities per unit energy, $N_{\mathrm{tp}}$, capture enough carriers in the etched region to establish a depletion region between the contacts, giving rise to a transport barrier whose height is determined by the pinned Fermi level. Conversely, when the total trap density is insufficient to compensate the charge associated with the implanted dopants, the Fermi level remains close to the valence band, preventing the formation of a significant depletion region and transport barrier (Figure 2c,d).

To quantify this effect, using TCAD simulations we extracted the activation energy from the temperature dependence of the low-bias resistance for different interface trap densities (Figure 2e). For low trap densities, the available interface states become saturated and are unable to significantly deplete the free-carrier concentration. Transport is not thermally activated because the Fermi level remains close to the valence band, allowing carriers to diffuse and drift across the device with little impediment. As $N_{\mathrm{tp}}$ increases, the trapped charge progressively shifts the Fermi level away from the valence-band edge in the etched region, increasing both the barrier height and the corresponding activation energy. At sufficiently high trap densities, the activation energy converges to $\approx 0.4$ eV, indicating strong Fermi-level pinning near the peak of the donor-like trap distribution ($E_{\mathrm{v}} + 0.4$ eV), beyond which further increases in trap density have little influence on transport (Figure 2f).

The transition from dopant-dominated to trap-dominated behavior occurs when the total interface trap density (for the donor-like trap distribution in Figure S7)

$$N_{\mathrm{t}} = \int_{E_{\mathrm{v}}}^{E_{\mathrm{c}}} g_{\mathrm{t}}(E_{\mathrm{t}})\, dE_{\mathrm{t}}\,, \tag{3}$$

becomes comparable to the sheet ionized acceptor density

$$N_{\mathrm{A,sheet}}^{-} = \int_{d_{\mathrm{etch}}}^{\infty} C(z)\, dz\,, \tag{4}$$

a quantity that represents the effective surface charge density due to the implanted boron ions, where the concentration profile $C(z)$ is taken from the etched surface. The etch depth, $d_{\mathrm{etch}} =$ 70 nm, is used in all TCAD simulations considered here. $N_{\mathrm{A,sheet}}^{-}$ is then $3.2 \times 10^{11}$ cm$^{-2}$, from which it is determined that $N_{\mathrm{tp}} = 1.5 \times 10^{12}$ cm$^{-2}$ eV$^{-1}$ is the critical interface-trap density per unit energy at the peak of the Gaussian trap distribution at which transport changes from dopant-dominated Ohmic conduction to nonlinear, thermally activated behavior. Consistent with this picture, the room-temperature *I–V* characteristics evolve from linear to strongly nonlinear as $N_{\mathrm{tp}}$ increases (Figure 2f). At lower trap densities ($N_{\mathrm{tp}} = 1.0 \times 10^{12}$ cm$^{-2}$ eV$^{-1}$), the device may remain nearly Ohmic at room temperature but become strongly nonlinear at 77 K, as in the cryogenic RNPUs of Ref. [10], whereas for $N_{\mathrm{tp}} \sim 1.0 \times 10^{14}$ cm$^{-2}$ eV$^{-1}$ pronounced nonlinearity is already present at room temperature.

These findings also explain the strong influence of surface treatment on device operation. HF etching followed by annealing reduces the interface-trap density, resulting in Ohmic room-temperature transport; nonlinear behavior can then only be recovered at lower temperatures, where thermal carrier generation is suppressed. In contrast, retaining the fluorocarbon layer preserves a high trap density, producing Fermi-level pinning approximately 300–400 meV above the valence band and maintaining a sufficiently low free-carrier concentration for nonlinear transport at room temperature. These results demonstrate that RNPU operation is governed by the balance between free carriers and trapped/depleted carriers, which together determine the emergence of the transport barrier and the resulting nonlinear behavior.

In contrast to boron RNPUs, arsenic RNPUs exhibit a stronger dependence on the etch depth (**Figure S9**). We attribute this difference to the nature of $P_{b1}$ centers at the silicon/oxide interface, which lie slightly below mid-gap, closer to the valence band edge [30]. The higher concentration of donor-like trap states leads to a stronger Fermi level pinning for *p*-type, hole-transport devices, explaining the closely packed activation energies.

## 2.3. Non-equilibrium carrier transport in RNPUs

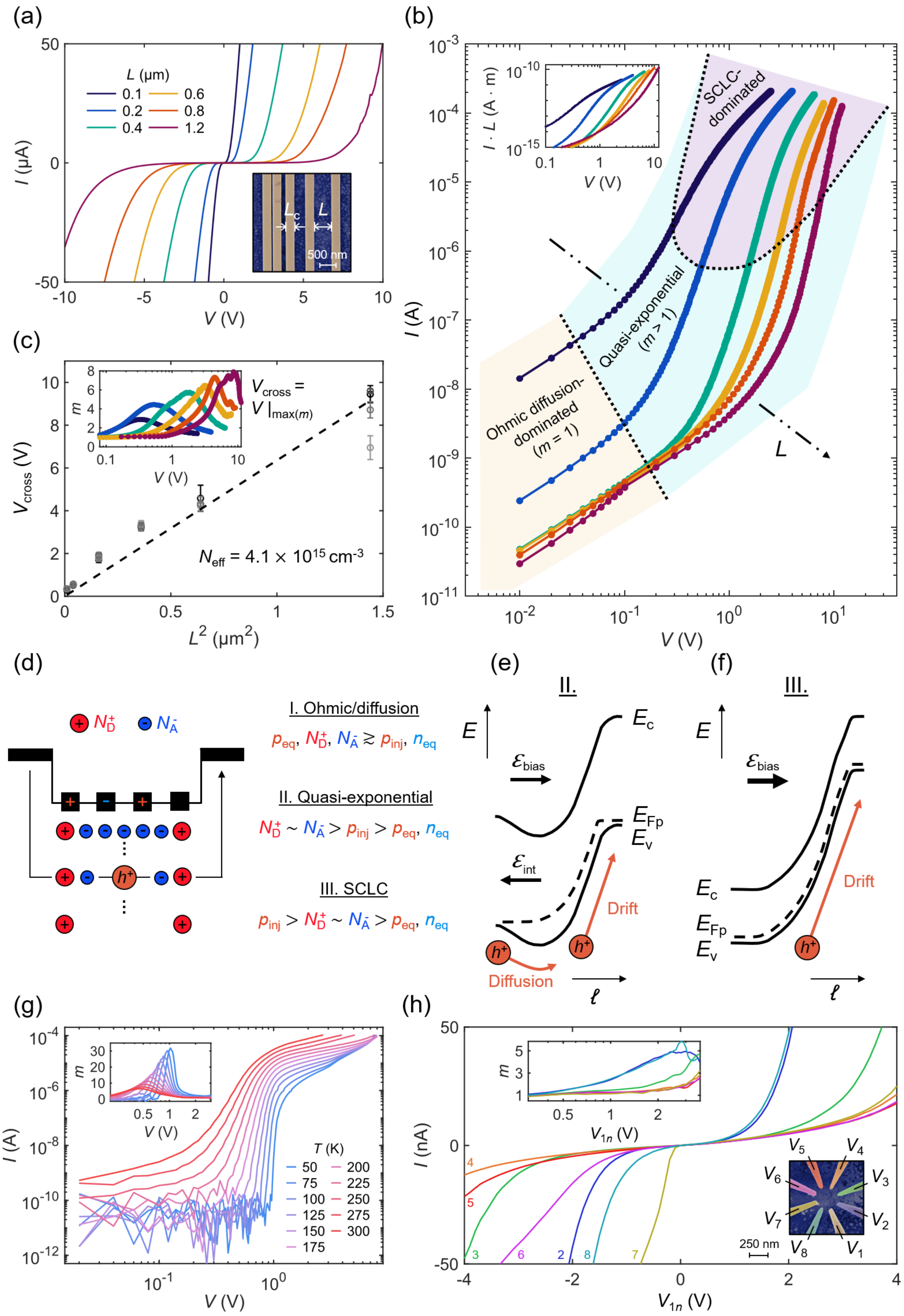

**Figure 3**. Experimental $I$–$V$ characteristics of boron-implanted (a)–(c),(g) transmission line model (TLM) structures and (h) RNPU. (a) $I$–$V$ characteristics for increasing $L$ with AFM image in inset indicating channel (contact) length $L$ ($L_c$). (b) Log–log plot of $I$–$V$ characteristics (etch depth = 70 nm, $L_c$ = 200 nm) with three characteristic regimes indicated. Inset: $I$–$V$ characteristics normalized to $L$. (c) Crossover voltages $V_{cross}$ versus $L^2$ for multiple structures (etch depth = 70 nm, $L_c$ = 200 nm). Inset: Local power-law exponent $m$ versus $V$ for various $L$. $V_{cross}$ corresponds to $V|_{max(m)}$, i.e., the voltage at which $m$ is maximal. (d) Schematic of relevant charge contributions governing hole transport, with hole current path highlighted. The charge-balance conditions in the three transport regimes are indicated as inequalities. (e-f) Schematic band diagrams along path $\ell$ in the quasi-exponential and SCLC-dominated transport regimes, respectively. (g) Log–log plot of $I$–$V$ characteristics at various $T$ ($L$ = 400 nm, $L_c$ = 500 nm). Inset: $m$ versus $V$ for various $T$. (h) Two-terminal $I$–$V$ characteristics of an RNPU. Current measured from biased electrode $V_1$ to grounded electrode $V_n$ ($n = 2, \dots, 8$), all other electrodes are kept floating. Top-left inset: Local power-law exponent $m$ versus $V$ for different electrode pairs. Bottom-right inset: False-colored AFM image of an RNPU.

In the following, we focus on hole-transport devices, i.e., boron RNPUs. While the high activation energy establishes thermally activated, Ohmic transport at low bias, it does not account for the pronounced nonlinearity observed at higher bias. We therefore investigate the current–voltage characteristics over an extended bias range using dedicated test structures. Specifically, bar-like transmission line model (TLM) structures [35] are fabricated on the implantation window, with the channel length $L$ between the bars systematically varied (**Figure 3**a). The characteristic nonlinearity is observed for all channel lengths, with its onset shifting to higher voltages for longer channels, indicating a transport mechanism governed by the electric field. To identify the underlying transport mechanism, the $I$–$V$ characteristics are analyzed on a log–log scale (Figure 3b), capturing all regimes from low to high bias. The degree of nonlinearity is quantified by the power-law exponent $m$ (Figure 3c), defined as

$$m = \frac{d\log(I)}{d\log(V)}. \quad (5)$$

Following the Ohmic ($m = 1$) low-bias regime, dominated by diffusion of dopant-provided free carriers not trapped by surface states, transport becomes strongly superlinear ($m > 1$). As carriers are injected, the potential barrier between the contacts is progressively lowered. Under these non-equilibrium conditions, transport in the active region can be described using Boltzmann (non-degenerate) statistics, such that the valence band edge $E_v(\ell)$, and hence the barrier, shifts with the electrostatic potential according to $E_v(\ell) = E_{v0} - |e|\psi(\ell)$, where $E_{v0}$ is its value at equilibrium electrostatic potential, $|e|$ is the elementary charge, and $\psi(\ell)$ is the position-dependent electrostatic potential inside the device, determined by the applied voltages and the space-charge distribution. The hole density is described by the quasi-Fermi level $E_{Fp}$, representing the effective chemical potential of holes under bias according to [34]:

$$p(\ell) = N_v \exp\left(-\frac{E_{Fp} - E_v(\ell)}{k_B T}\right), \quad (6)$$

with $N_v$ the effective density of states in the valence band.

The $I$–$V$ characteristics in the intermediate regime exhibit a strongly curved increase on a log–log scale (Figure 3b), which suggests an exponential dependence [36], as would arise from the Boltzmann statistics of Equation (6) under diffusion-limited transport. Further analysis (**Figure S10**) shows that transport cannot be described by purely a diffusion current though. This indicates that the implicit assumption of a linear relation between the electrostatic potential $\psi(\ell)$ and the applied bias voltage $V$ is not valid. The energy bands are nonetheless strongly modulated by $V$. TCAD simulations show that $E_{Fp} - E_v(\ell)$ depends nonlinearly on $V$ (**Figure S11**), demonstrating that the apparent exponential behavior does not arise from a pure diffusion current, but from a nonlinear electrostatic response. Diffusion, however, still contributes significantly, as discussed below.

We considered several charge transport mechanisms to explain the nonlinear non-equilibrium behavior. In particular, for transport along interface states, both hopping (**Figure S12**) and Poole–Frenkel emission (**Figure S13**) were examined. However, the observed length scaling of the measured current, combined with TCAD results indicating that the dominant current path extends into the semiconductor volume rather than being confined to the interface, points to a bulk transport mechanism. Specifically, our analysis identifies space-charge-limited current (SCLC), which is known to dominate transport in $n^+nn^+$ and $p^+pp^+$ structures [37]. In planar geometries, the SCLC density $J_{\mathrm{SCLC}}$ follows a modified Mott–Gurney law with a weaker length dependence than the $L^{-3}$ dependence of the classical sandwich configuration [38–41]:

$$J_{\mathrm{SCLC}} \propto \mu\epsilon\frac{V^2}{L^2}, \tag{7}$$

where $\mu$ is the mobility and $\epsilon$ is the permittivity.

Moreover, at high electric fields, where velocity saturation sets in, the mobility in silicon becomes field-dependent, often described by the Caughey–Thomas mobility model [42]:

$$\mu(\mathcal{E}) = \mu_0\left[\frac{1}{1+\left(\frac{\mu_0\mathcal{E}}{v_{\mathrm{sat}}}\right)^{\alpha}}\right]^{\frac{1}{\alpha}}, \tag{8}$$

with $\mu_0$ the low-field mobility, $\mathcal{E}$ the magnitude of the lateral electric field (parallel to the current), $v_{\mathrm{sat}}$ the saturation velocity and $\alpha$ a material-dependent parameter on the order of 1. In the high-field limit, SCLC modified by velocity saturation (VS) alters both the voltage and length dependence of the current density [43]:

$$J_{\mathrm{SCLC-VS}} \propto v_{\mathrm{sat}}\epsilon\frac{V}{L}. \tag{9}$$

The inset of Figure 3b indicates an approximate $L^{-1}$ scaling of the current at high bias, which is consistent with SCLC in the velocity-saturation regime. In addition, the slope plots in the inset of Figure 3c show that, in the high-field limit, the voltage dependence approaches Ohmic behavior ($m \approx 1$) for shorter channels. These findings are further supported by TCAD simulations: including field-dependent mobility reproduces the observed channel-length and voltage scaling, consistent with Equations (7) and (9) (**Figure S14–S16**). We thus find three distinct transport regimes: a low-bias, diffusion-dominated Ohmic regime ($m = 1$), a strongly nonlinear quasi-exponential regime ($m > 1$), and an SCLC-dominated regime, modified by velocity saturation at high fields ($m \approx 1$). In the latter two regimes, transport is governed by space charge.

The presence of a space-charge effect is supported by considering the charge contributions in the Poisson equation. The space-charge effect occurs when the injected hole density exceeds both the equilibrium carrier density and the doping contributions [34]. Considering the charge density in a bulk region of the device, far from the contacts and assuming predominantly planar current flow, the relevant contributions are given by

$$\frac{\epsilon}{|e|}\frac{d\mathcal{E}}{d\ell} = \left[\left(p(\ell) - p_{\mathrm{eq}}\right) - \left(n(\ell) - n_{\mathrm{eq}}\right) + N_{\mathrm{D}}^{+} - N_{\mathrm{A}}^{-}\right], \tag{10}$$

where $p_{\mathrm{eq}}$ ($n_{\mathrm{eq}}$) is the equilibrium hole (electron) concentration, assumed constant in this volume, and $p(\ell) = p_{\mathrm{inj}}(\ell) + p_{\mathrm{eq}}$ and $n(\ell) = n_{\mathrm{inj}}(\ell) + n_{\mathrm{eq}}$ denote the total hole and electron densities, respectively, with $p_{\mathrm{inj}}$ and $n_{\mathrm{inj}}$ the concentration of the injected carriers. $N_{\mathrm{D}}^{+}$ is the ionized donor concentration due to the phosphorus background doping, and $N_{\mathrm{A}}^{-}$ is the ionized acceptor concentration due to the implanted boron dopants. Figure 3d schematically shows these contributions. In a hole-transport device, both bulk and injected electron densities are negligible. Moreover, because interface states capture most free carriers near the central surface region, charge transport shifts into the bulk, such that $N_\mathrm{A}^- \sim N_\mathrm{D}^+$ (Section 2.4). Because the degenerately doped contacts efficiently inject holes into the bulk, while interface traps strongly suppress the equilibrium carrier density in the etched region, the dominant competition is between $p_\mathrm{inj}(\ell)$, and the fixed ionized acceptors $N_\mathrm{A}^-$ and donors $N_\mathrm{D}^+$. The equilibrium hole density is relevant only in the low-bias Ohmic regime, and is rapidly dominated by injected carriers at relatively higher biasing. By suppressing the equilibrium carrier concentration, interface states enable space-charge effects and the resulting SCLC behavior responsible for the observed nonlinear transport.

The injected holes are driven by the lateral electric field $\mathcal{E}_\mathrm{bias}$ arising from the externally applied bias voltage $V$. Opposing this is an internal electric field $\mathcal{E}_\mathrm{int}$, associated with the depletion-induced transport barrier which counteracts the injection of holes. In this approximately exponential transport regime, the depletion regions formed near the surface and at the *p-n* junction give rise to pronounced band bending (Figure 3e, S11), steep carrier concentration gradients, and a strongly nonlinear lateral electric-field profile. The diffusion current is therefore expected to contribute significantly to charge transport [36], further enhanced by the boron concentration gradient below the contacts. As the injected hole concentration increases and exceeds the net doping contribution, the injected carriers become the dominant contribution to the space charge. As a result, the externally applied field overwhelms the internal field, causing the transport barrier to collapse. Provided that the diffusion current is negligible compared to the drift current, transport then enters the space-charge-limited-current regime (Figure 3f) [34].

We define a crossover voltage $V_\mathrm{cross}$ separating the quasi-exponential and SCLC regimes, corresponding to the condition $p > \left|N_\mathrm{A}^- - N_\mathrm{D}^+\right|$ across the device, analogous to the trap-filled-limit voltage in insulators with a bulk trap density [40]:

$$V_{\mathrm{cross}} = \frac{\sigma_{\mathrm{eff}}}{C_0} = \frac{eN_{\mathrm{eff}}L^2}{\epsilon} \approx \frac{e\left|N_{\mathrm{A}}^{-} - N_{\mathrm{D}}^{+}\right|L^2}{\epsilon}, \tag{11}$$

where $\sigma_{\mathrm{eff}}$ is the effective injected sheet charge required to satisfy $p > \left|N_{\mathrm{A}}^{-} - N_{\mathrm{D}}^{+}\right|$, $C_0$ the areal capacitance between the electrodes, and $\epsilon = 11.8\ \epsilon_0$ [22] the permittivity of silicon. Here, $N_{\mathrm{eff}}$ denotes an effective carrier density, which approximates a net doping concentration. Additional carriers must be injected to overcome the hole density along the doping gradient beneath the contacts, as well as to compensate thermally generated carriers captured by interface traps in the etched region. As a result, the value of $N_{\mathrm{eff}}$ extracted from the crossover voltage is expected to overestimate $\left|N_{\mathrm{A}}^{-} - N_{\mathrm{D}}^{+}\right|$. $V_{\mathrm{cross}}$ thus corresponds to the bias voltage at which the total injected charge exceeds the fixed and trapped charge in the device, such that the electrostatics solely becomes governed by injected carriers, marking the onset of space-charge-limited transport. In order to extract $V_{\mathrm{cross}}$, we approximate that the peak voltage in the plot of $m(V)$ (Figure 3c) is the voltage at which this onset occurs (Section 4.8). Fitting Equation (11) to the crossover voltages extracted from four test structures yields $N_{\mathrm{eff}} = 4.1 \times 10^{15}\ \mathrm{cm}^{-3}$, three orders of magnitude lower than expected for an etch depth of ~70 nm under the assumption $N_{\mathrm{eff}} = \left|N_{\mathrm{A}}^{-} - N_{\mathrm{D}}^{+}\right|$, based on the SIMS profile in Figure S2. The extracted value is instead much closer to the background phosphorus doping inferred from the substrate resistivity specified by the wafer supplier (Section 4.1), which lies in the range $5 \times 10^{14}\ \mathrm{cm}^{-3}$ and $5 \times 10^{15}\ \mathrm{cm}^{-3}$. In Section 2.4, we investigate the respective roles of the two doping contributions and show that the background doping is essential for the emergence of the quasi-exponential regime.

With the dominant charge transport mechanism identified, we revisit the temperature dependence of the *I–V* characteristics for a 400 nm channel. As shown in Figure 3g, $V_{\mathrm{cross}}$ increases with decreasing temperature. This behavior can be understood from the reduction in $p_{\mathrm{eq}}$, which is governed by thermally activated generation of holes. As the temperature decreases, fewer holes are thermally excited, leading to a lower $p_{\mathrm{eq}}$. In addition, the high density of surface trap states within the band gap further suppresses hole activity, as trapped holes are less likely to be thermally re-emitted at lower temperatures. As a result, the balance between $p(V) = p_{\mathrm{eq}} + p_{\mathrm{inj}}(V)$ and $\left|N_{\mathrm{A}}^{-} - N_{\mathrm{D}}^{+}\right|$ becomes temperature-dependent: a lower $p_{\mathrm{eq}}$ requires a larger injected carrier density, which in turn necessitates a higher applied voltage. This leads to an upward shift of $V_{\mathrm{cross}}$, consistent with the observed temperature dependence. The pronounced temperature dependence of the current in the quasi-exponential regime is consistent with established behavior [36, 41] and indicates that hole diffusion contributes significantly to transport in this regime. On the other hand, the weak temperature dependence in the high-field regime is consistent with SCLC, where residual temperature dependence primarily arises from the carrier mobility [36].

Analogously to our analysis for boron RNPUs, we apply the same procedure to arsenic RNPUs and obtain $N_{\mathrm{eff}} = 3.5 \times 10^{15}\ \mathrm{cm}^{-3}$ (**Figure S17**). In this case, the background boron doping contributes the ionized acceptors $N_{\mathrm{A}}^{-}$, while the implanted arsenic dopants contribute the ionized donor concentration $N_{\mathrm{D}}^{+}$. We further note that the contact geometry influences the onset of the space-charge effect (**Figure S18**). In particular, increasing $L_{\mathrm{c}}$ increases the availability of equilibrium free carriers, thereby reducing the crossover voltage and leading to an earlier onset of the SCLC regime (Figure 3c,e). In RNPUs, due to the contacts becoming increasingly narrow as the active region of the device is reached ($L_{\mathrm{c}}$ ~ 20 nm), $V_{\mathrm{cross}}$ is higher compared to the test structures investigated here (Figure 3h). Within the measured voltage range, the crossover voltage is first reached by electrodes in close proximity, further emphasizing the channel-length dependence of space-charge phenomena. The voltage range relevant for RNPU operation lies entirely within the quasi-exponential regime ($< V_{\mathrm{cross}}$), with $V_{\mathrm{cross}}$ (and the current) varying as a function of both voltage polarity and electrode separation.

Under negative bias, holes are injected from the grounded electrode, resulting in a redistribution of positive space charge between the electrodes that differs from the case of positive bias. In addition, slight local variations in doping and trap density, as well as fabrication-induced differences between contacts, can further influence the *I*–*V* characteristics. These observations underscore the additional complexity of charge transport in RNPUs arising from their electrode configuration and geometry. While the multi-terminal geometry of RNPUs introduces additional electrostatic complexity (to be discussed in Ref. [28]), we argue that the underlying charge transport mechanism is the same.

### 2.4. TCAD analysis of background doping in RNPUs

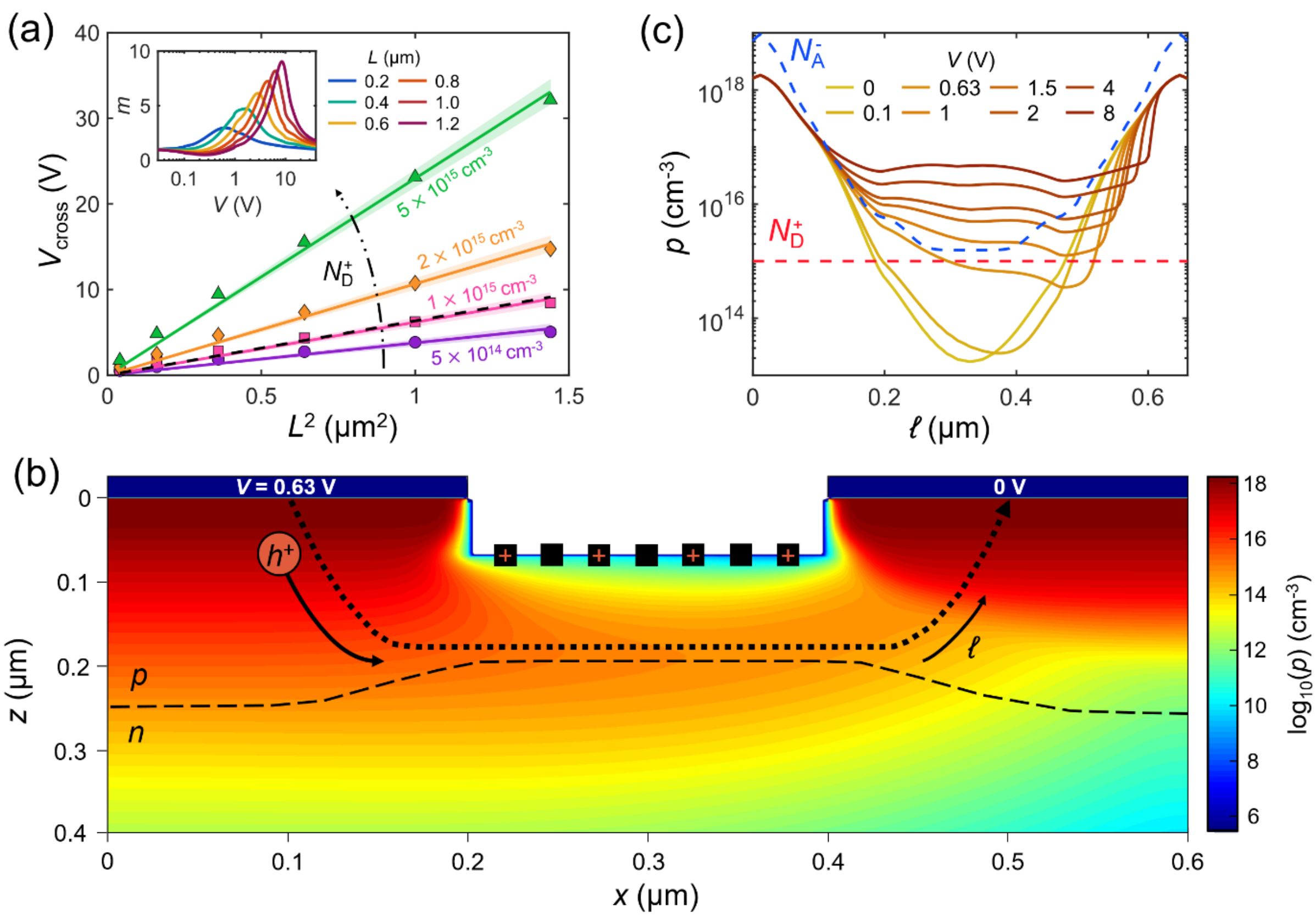

**Figure 4**. TCAD device simulations of *p*-type TLM structure, highlighting the role of background doping and traps. (a) $V_{cross}$ versus $L^2$ for different ionized donor doping concentrations $N_D^+$ (dot-dashed arrow). Shaded regions indicate fit uncertainty; dashed line shows experimental fit from Figure 3c.

Inset: Local power-law exponent $m(V)$ for varying channel lengths ($N_D^+ = 10^{15}$ cm$^{-3}$). (b) Contour plot of free hole concentration $p$ with bias voltage $V = 0.63$ V $\approx V_{cross}$. Traps at the Si/SiO$_x$ interface are represented by black squares. Hole current flows along path $\ell$. Dashed line indicates the position of the *p*-*n* junction. (c) Hole concentration along path $\ell$ in (b) for increasing $V$. Red and blue dashed curves indicate $N_D^+$ and $N_A^-$, respectively.

Silvaco TCAD device simulations [22] support the existence of three characteristic regimes, confirm the presence of space-charge effects, and demonstrate the key role of background dopants in the nonlinear transport mechanism. Employing a two-dimensional simulation geometry with varying channel length $L$, similar $\propto L^2$ scaling is obtained for different background doping concentrations (**Figure 4**a). The simulations also indicate that $N_{eff} \sim |N_A^- - N_D^+|$: for $N_D^+ = 1.0 \times 10^{15}$ cm$^{-3}$, we obtain $N_{eff} = 4.0 \times 10^{15}$ cm$^{-3}$, which is comparable to the experimentally extracted value. The background doping is essential to device functionality, as the depletion region associated with the *p*-*n* junction shapes the internal electric-field profile and thereby governs the onset of tunable nonlinearity with applied bias. **Figure S19** shows that increasing the background doping shifts the onset of the SCLC regime to higher voltages, while nonlinearity is strongly suppressed when the background dopants are of the same type as the implanted dopants. In the latter case, the background doping increases the equilibrium hole concentration, and the *p*-*n* junction with the substrate is effectively absent. Without this junction, the bulk depletion region does not form, leading to a higher hole concentration and weaker band bending associated with $\mathscr{E}_{int}$, which lowers the transport barrier for holes and suppresses the diffusion current contribution, effectively bypassing the quasi-exponential regime. Strong nonlinearity therefore relies on depletion effects, which require the presence of dopants of opposite polarity to the injected carriers.

Figure 4b shows a contour plot of the free hole concentration at a voltage approximately equal to $V_{\text{cross}} \sim 0.63$ V, illustrating that the current flows in the bulk, several tens of nanometers beneath the silicon surface. The interface-trap-induced depletion region suppresses the free carrier concentration near the etched surface, forcing current flow deeper into the device, in the vicinity of the *p-n* junction formed with the substrate. The Ohmic contacts and gradient in $N_{\text{A}}^{-}$ ensure efficient carrier injection and a low-resistance path to the channel. As the injected hole density increases, the conductive region expands while the depletion region is pushed further into the bulk to maintain charge balance. Figure 4c depicts the hole concentration *p* along the cross-sectional path $\ell$ for different bias voltages, indicating that once *p* exceeds $|N_{\text{A}}^{-} - N_{\text{D}}^{+}|$ throughout the device (above ~$V_{\text{cross}}$), space charge becomes injection-dominated, and transport enters the space-charge-limited regime. Below ~$V_{\text{cross}}$, steeper *p* gradients indicate a significant contribution of carrier diffusion to the current density, and hence to the device nonlinearity. Because the hole current penetrates slightly deeper into the semiconductor, the ionized acceptor concentration $N_{\text{A}}^{-}$ along this path becomes comparable to, and in some regions smaller than, *p* and $N_{\text{D}}^{+}$. The role of the ionized acceptors has thus far not been discussed in detail. If the boron SIMS profile is directly assumed to represent $N_{\text{A}}^{-}$, TCAD simulations do not reproduce the experimentally observed nonlinearity (**Figure S20**), instead yielding sub-Ohmic *I–V* characteristics, likely due to mobility degradation from impurity scattering [22]. Reproducing the experimental behavior requires partial dopant deactivation within the channel. Reactive-ion etching is known to electrically deactivate dopants through hydrogen diffusion and the formation of electrically inactive dopant complexes [44–46], particularly in hydrogen-containing plasmas [44] such as the $CHF_3$-based plasma used here. Low-temperature annealing near 450°C has been shown to restore dopant activity [45], suggesting that the annealing performed as part of the RNPU device treatment may simultaneously passivate $P_{\text{b}}$ centers while reactivating electrically inactive dopants. More generally, independent of the underlying mechanism, the emergence of the quasi-exponential regime requires a sufficiently low equilibrium hole concentration in the channel region to enable strong band bending. Both interface traps and electrically inactive (acceptor) dopants contribute to this reduction in carrier density, thereby enabling nonlinear space-charge effects.

As discussed in Section 2.1, treated boron devices have been shown to be functional at room temperature by applying a positive potential to the *n*-type substrate [10]. We can now explain this in the framework that we have presented. A positive substrate potential drives holes away from the bulk of the device toward the contacts which are at a lower potential, widening the depletion region at the *p*-*n* junction, thereby reducing $p_{\mathrm{eq}}$ in the bulk (**Figure S21**). This reduction enhances carrier concentration gradients and, consequently, the diffusion current, while shifting the onset of the SCLC regime to higher voltages. As a result, the slope $m$ increases within the same voltage range, which is favorable for room-temperature operation. These findings further indicate that the substrate potential can serve as an additional tuning parameter for untreated, room-temperature RNPUs, particularly for controlling the quasi-exponential regime prior to the onset of SCLC.

Notably, the voltage at which the injected hole concentration exceeds $\left|N_{\mathrm{A}}^{-} - N_{\mathrm{D}}^{+}\right|$ across the device corresponds closely to $V_{\mathrm{cross}} = V|_{\max(m)}$ (Figure S11), but since it is based on a simplified 1D approximation, it may help explain deviations of the extracted $V_{\mathrm{cross}}$ from the ideal behavior given in Equation (11). Deviation between simulated and experimental crossover voltages may arise from a limitation of the two-terminal TCAD model, which does not fully capture the influence of adjacent electrodes in the TLM geometry of Figure 3a. In the experimental structure, nearby contacts inject holes that extend into the channel region, increasing the local equilibrium hole density. As a result, a lower applied voltage (and hence fewer injected carriers) is required to reach the condition $p > \left|N_{\mathrm{A}}^{-} - N_{\mathrm{D}}^{+}\right|$. Both experimental

and simulated $V_{\mathrm{cross}}$ versus $L^2$ trends show that Equation (11) underestimates $V_{\mathrm{cross}}$, particularly for shorter channel lengths. This deviation likely results from the contact lengths becoming comparable to the channel length, thereby violating the assumptions of the ideal electrostatic model in Ref. [38]. Moreover, due to the bulk nature of the current flow, defining a particular voltage at which the injected carriers have overcome the equilibrium hole concentration and dopant concentrations is challenging. Figure 4b and 4c highlight this by showing that the space-charge distribution, and thus the electric field profile, varies non-uniformly across the device.

## 3. Conclusions

Using both boron and arsenic RNPU devices together with two-terminal TLM structures that capture the relevant transport mechanism, we establish that space-charge effects govern the charge transport underlying RNPU functionality in the devices studied here and in Ref. [6]. The modified fabrication route employed here, in which a previously used treatment step is omitted, enables robust room-temperature operation. This shows that disorder is not essential for their operation. Instead, the behavior is governed by two coupled charge balances. First, the interplay between RIE-induced traps at the silicon/silicon oxide interface and implanted dopants sets the equilibrium carrier concentration. Sufficient suppression of this carrier density is required to enable tunable nonlinearity at room temperature, a condition that may be further enhanced by dopant deactivation. Second, once this condition is met, the competition between injected carriers and ionized dopants determines the dominant source of space charge. In general, a sufficiently low equilibrium carrier density in the active region allows strong band bending and the formation of an internal electric field (and associated transport barrier). In the present devices, the presence of oppositely doped regions provides a depletion region that further suppresses the hole concentration and establishes a spatially varying electric field, enabling the nonlinear and tunable response underlying device functionality.

While superquadratic SCLC behavior is typically associated with trap-dense materials such as organics and perovskites [47–51], we show that RNPUs constitute an analogous system in crystalline silicon. Crucially, in contrast to these materials, the nonlinearity is governed primarily by dopants rather than traps. The observed $L^2$ scaling of the crossover voltage is analogous to the trap-filled-limit voltage, but arises from a fundamentally different physical origin. Because the mechanism is based on drift-diffusion and electrostatics in an ordered system, device behavior is expected to be more predictable and reproducible, facilitating more consistent and transferable training across devices.

These insights provide a route toward planar silicon implementations with engineered doping profiles, reducing the reliance on etched structures and interface traps while strengthening compatibility with CMOS fabrication. Together with TCAD-based modeling, this establishes a physics-driven framework for optimizing RNPU devices and reducing device-to-device variation.

## 4. Methods

### 4.1. Device fabrication

Fabrication of RNPUs follows the process flow described in Ref. [10]. After 26 μm × 60 μm implantation windows have been defined by wet etching into a thermally oxidized (100) silicon wafer (approximately 280 nm thick oxide), the exposed silicon is again oxidized to provide a stopping layer to ion implantation (25 and 35 nm thick for the As and B implants, respectively). In this way, the As and B peak concentrations are closer to the silicon surface. For the B implantation, the ion energy is 9 keV and the dose is $3.5 \times 10^{14}$ at/cm$^2$, while for the As implantation the ion energy is 35 keV and the dose $1.0 \times 10^{14}$ at/cm$^2$. The B implantation is done on lightly phosphorus-doped $n$-type wafers with resistivities ranging between 1 and 10 Ω·cm, whereas the As implantation is done on lightly B-doped $p$-type wafers with resistivities ranging between 5 and 10 Ω·cm, as given by the supplier Okmetic. After 7 s rapid-thermal annealing at 1,050˚C and 1,100˚C for the boron- and arsenic-implanted wafers, respectively, the stopping layer is removed, markers are defined and the wafers are diced into $1 \times 1$ cm$^2$ samples. Both RNPUs and TLM structures are fabricated by means of electron-beam lithography, electron-beam evaporation and a lift-off process (Figure S1a,b). The metal structures consist of a ~1.5 nm/25 nm Ti/Pd stack.

The implanted dopant profile, as determined by secondary-ion mass spectroscopy (SIMS), is shown in Figure S2 for both B and As implants following rapid thermal annealing (RTA). It is possible that a fraction of dopants are inactive following the RTA step, though it is reported in Ref. [52] that for $T \geq$ 1,000˚C all boron dopants are activated in ~10 s for relevant doses. It is well-known that dry etching can result in a severe reduction of dopant activity, and in some cases even complete deactivation, primarily attributed to hydrogen passivation of dopant impurities which forms electrically inactive complexes [44–46]. Besides the surface states generated in the band gap by $P_b$ centers, the dopant inactivity may thus also contribute to a significant reduction of the free carrier concentration. In our TCAD simulations, we assume that the fractions of active dopants are 20% below the contacts, and 5% in the channel region that is directly exposed to the plasma. We evaluate the sensitivity of the TCAD simulations to the fraction of inactive dopants, both below the contacts and in the channel regions (Figure S20).

The critical reactive-ion etching (RIE) step is performed in a custom-built parallel-plate RIE reactor. We make use of a 5:1 $CHF_3/O_2$ mixture at a pressure of 25 mTorr, a substrate temperature fixed at 10˚C, and an RF power equal to 25 W, driven at 13.56 MHz. This produces $C_{\alpha}H_{\beta}F_{\gamma}$ and F radicals to etch silicon (dioxide), with the $C_{\alpha}H_{\beta}F_{\gamma}$ radicals responsible for polymerizing the surface [53], in particular for mixtures containing fluorocarbon gases with $\gamma/\alpha < 3$, which includes $CHF_3$ [54]. The polymer layer and sufficient oxygen ion bombardment lead to an anisotropic etching profile (Figure S3). After RIE, a thin polymer layer remains on the etched surface, providing partial protection to the silicon surface. Since the etch depth is determined via atomic-force microscopy (AFM), it is thus underestimated unless the fluorocarbon layer is fully removed, e.g. by means of HF wet etching. This was part of the treatment followed in previous "treated" devices [10].

Prior to RIE, the step height is determined, yielding the thickness of the metallic contacts (Figure S4a). Following RIE (etch time = 3 min), the step height is measured again, from which an apparent etch depth of $40.8 \pm 3.8$ nm is extracted for one of the four TLM structures analyzed in Figure 3c. This apparent etch depth also yields optimal RNPU performance, providing strong nonlinearity while maintaining sufficient room-temperature conductivity to enable Boolean logic gates with high fitness, i.e., clear separability between high and low states. The RNPUs in Figure 2a correspond to apparent etch depths of ~25, ~45, and ~75 nm for etch times of 1.5, 3, and 4.5 min, respectively.

Following HF etching, we found that the real etch depth of the TLM structure is $70.3 \pm 9.6$ nm due to a $29.5$ nm $\pm$ $10.8$ nm thick fluorocarbon layer. This is also the etch depth used in our TCAD simulations, as it more accurately reflects the peak doping level in the silicon. Similar observations can be made on RNPUs (Figure S4b), however, due to the difference in geometry, the etch rate is reduced and the deposited fluorocarbon layer is also slightly thinner. This is explained by the smaller feature sizes and correspondingly less silicon area in the active area of RNPUs, which affects how the radicals interact with the surface. Besides geometrical effects, the etch rates of boron- and arsenic-implanted silicon also differ due to band bending effects at the surface [55]. The etch rate of arsenic-implanted devices is higher than that of the boron-implanted devices since the uncompensated $As^+$ donors attract negatively charged $F^-$ ions, enhancing the etch rate. In contrast, the etch rate is reduced by means of Coulombic repulsion for a silicon lattice containing $B^-$ acceptor ions.

### 4.2. Influence of HF wet etch on electrical properties

The influence of the HF wet etch on the electrical properties of the device is an order-of-magnitude conductance reduction and a strong suppression of the nonlinearity for the measured voltage range (Figure S5). The electrically active carrier concentration appears to be substantially reduced, possibly due to exposure of the silicon surface, which increases the density of dangling-bond-related surface states and enhances carrier trapping. This shifts the onset of strong nonlinearity to higher voltages, where device breakdown may occur before any functionality is attained. For an etch depth of 70 nm, and ~70 nm electrode spacing (adjacent electrodes), the breakdown voltage is estimated to be ~5.3 V, using the critical breakdown field of silicon of $2.5 \times 10^5$ V/cm [34]. Given the gradient in the doping and the influence of interface states on the free carrier concentration, space-charge effects – and the resulting nonuniform electric field – cannot be neglected, and may significantly alter the critical breakdown field of the device. Nevertheless, the device resistance exceeding giga-ohms, the resulting low signal-to-noise ratio, and the low degree of nonlinearity for typical voltages have consequences for both readout and device functionality. This necessitates annealing in order to lower the resistance, however, as reported in Ref. [10] and in Figure 1c, this in turn requires lowering the temperature to suppress the free carrier concentration such that the strong nonlinearity induced by space-charge effects can re-emerge. These experimental observations are supported by TCAD simulations, which show that for a given trap density the *I*–*V* characteristics are linear at room temperature and become nonlinear at 77 K (Figure 3f).

### 4.3. Electrical characterization methods

*I–V* characteristics at room temperature are measured in an MPI TS300-SE Probe System with a Keithley 4200A-SCS Parameter Analyzer. The substrate, which lies on the wafer chuck, is left floating for all measurements concerned here. One electrode remains grounded as the potential on the other electrode is varied from -*V* to *V*. For this reason, the log–log data that is plotted belongs to the positive range of the sweep since the current has adequately settled. Follow-up measurements were performed to ensure consistency.

The temperature-dependent measurements in Figure 2 are performed in a Suss MicroTec PM300 Probe System with a cooling unit, flushed with nitrogen under vacuum conditions. For the temperature-dependent measurements that are plotted in Figure 1c, 3g, S9 and S12, we use the PPMS Dynacool system which makes use of evaporated liquid helium. This allows for characterization under cryogenic conditions, which is relevant for treated devices which operated at 77 K, and investigating hopping phenomena.

### 4.4. Evaluation of device functionality

The solved Boolean logic gates in Figure 1d and S6 are obtained by applying step voltages to the input electrodes, generating 10,000 random control voltage configurations, and measuring the current at the output electrode. The input voltages are applied such that they represent a 2-bit logic circuit. They are applied to the contacts adjacent to the one opposite the output, and a set of control voltages capable of solving the desired task are found. Typically, a genetic algorithm is made use of in order to evolve the set of control voltages to obtain the highest possible fitness. The details of how this genetic algorithm is used for Boolean logic gate evolution is available in previous work [10, 56]. The output current $\bar{I}_{\text{out}}$ for each input sequence is fitted with

$$\bar{I}_{\text{out}} = m_{\text{BL}}\bar{X}_{\text{out}} + c_{\text{BL}}, \tag{12}$$

where $\bar{X}_{\text{out}}$ is the expected logic output (either 1 or 0), $m_{\text{BL}}$ is a measure of the separation of the logic high and low levels, and $c_{\text{BL}}$ is an offset term. The fitness of the Boolean logic gate is then evaluated by

$$F_{\text{BL}} = \frac{m_{\text{BL}}}{\sqrt{r_{\text{ss}}} + k_{\text{BL}}c_{\text{BL}}}, \tag{13}$$

where $r_{\text{ss}}$ is the fitting residual and $k_{\text{BL}}$ an empirical constant. We choose $k_{\text{BL}}$ to be 0.2 as in Ref. [10], and plot the Boolean logic gate waveforms with highest fitness for each of the major logic gates in Figure 1d and S6. In one run consisting of 10,000 configurations on the same device, it is possible to find configurations which use the same threshold (approximately 40 nA) for all gates, with the exception of the NAND gate.

### 4.5. Simulation methods

Simulations are performed using Victory TCAD from Silvaco, a standard tool used in the semiconductor industry. The simulation comprises Victory Process (Version 8.42.5.R) and Victory Device (Version 1.24.0.R), which concern the fabrication- and device physics-based aspects, respectively, of the simulation. For our purposes, we utilize Victory Process to generate a two-dimensional, two-terminal structure, as illustrated in Figure 4b. The simulations are thus a good approximation of the TLM test structure. This includes the assumption of a uniform background doping, as well as generating the mesh itself. The implantation profile of dopants opposite to the background doping is incorporated into the mesh using the concentration profile(s) shown in Figure S2. We then make use of Victory Mesh (Version 1.11.1.R) to reconstruct and transfer the structure to Victory Device. Alternatively, the profile can be implemented directly in Victory Device without Victory Process. The interface traps are assumed to be uniform along the entire silicon/silicon oxide interface. We assume $P_{b0}$ centers are the dominant contribution to the surface states in the bandgap, represented by two Gaussian distributions, one acceptor-like (in the vicinity of the conduction band) and one donor-like (in the vicinity of the valence band). The density of states $g_\mathrm{t}$ of such a Gaussian trap distribution is given by

$$g_\mathrm{t} = N_\mathrm{tp} \exp\left[-\left(\frac{E_\mathrm{t} - E_\mathrm{tp}}{W_\mathrm{t}}\right)^2\right], \tag{14}$$

where $N_\mathrm{tp}$ is the trap density per unit energy at the peak of the Gaussian, $E_\mathrm{tp}$ is the energy level at which the Gaussian peaks (relative to the respective band edge), $E_\mathrm{t}$ is the energy level of the trap state, and $W_\mathrm{t}$ is the characteristic decay energy of the Gaussian [22]. We assume two identical distributions, with $N_\mathrm{tp} = 2 \times 10^{14}\ \mathrm{cm}^{-2}\ \mathrm{eV}^{-1}$, $E_\mathrm{tp} = 400$ meV, and $W_\mathrm{t} = 120$ meV (Figure S7). The distributions were placed such that they were near mid-gap in accordance with Refs. [29, 30], with the actual density being chosen such that sufficient traps exist to pin the Fermi level (Figure 2e). We expect a density of interface states several orders of magnitude above typical interface state densities of high-quality interfaces ($D_\mathrm{it} \sim 10^{10} - 10^{11}\ \mathrm{cm}^{-2}\ \mathrm{eV}^{-1}$) [57] due to the unannealed, RIE-damaged surface.

The simulation is based on a few fundamental physical equations; the equations are given as described in Ref. [22]. Computations are performed using Maxwell-Boltzmann statistics, so that the probability of a state with energy $E$ being occupied is given by

$$f(E) = \frac{1}{1 + \exp\left(\frac{E - E_\mathrm{F}}{k_\mathrm{B}T}\right)} \approx \exp\left(-\frac{E - E_\mathrm{F}}{k_\mathrm{B}T}\right). \tag{15}$$

We found that when instead using Fermi-Dirac statistics, the computed $I$–$V$ characteristics were identical, supporting the assumption that $E - E_{\mathrm{F}} \gg k_{\mathrm{B}}T$ and thus justifying the use of Boltzmann statistics, which reduces computation time. The electrostatics in the device is described by Poisson's equation, with the device simulator considering five overall contributions [22]:

$$\nabla \cdot (\epsilon \mathcal{E}) = \rho = \left[|e|\left(p - n + N_{\mathrm{D}}^{+} - N_{\mathrm{A}}^{-}\right) + Q_{\mathrm{T}}\right], \tag{16}$$

where $\mathcal{E}$ is the electric field, $\rho$ is the bulk charge density, and $Q_{\mathrm{T}}$ is the charge density contribution due to traps. The contribution due to traps is computed based on the integral of the density of states and the trap occupancy:

$$Q_{\mathrm{T}} = |e| \int_{E_{\mathrm{v}}}^{E_{\mathrm{c}}} \left[g_{\mathrm{t,D}}(E_{\mathrm{t}}) \cdot f_{\mathrm{t,D}} - g_{\mathrm{t,A}}(E_{\mathrm{t}}) \cdot f_{\mathrm{t,A}}\right] dE_{\mathrm{t}}\,, \tag{17}$$

where we consider both donor-like and acceptor-like distributions, with $f_{\mathrm{t}}$ the probability of defect ionization. In steady-state equilibrium, for the donor-like states, this probability is given by

$$f_{\mathrm{t,D}} = \frac{\sigma_{\mathrm{n}} v_{\mathrm{n}} n_{\mathrm{t,D}} + \sigma_{\mathrm{p}} v_{\mathrm{p}} p}{\sigma_{\mathrm{n}} v_{\mathrm{n}}\left(n + n_{\mathrm{t,D}}\right) + \sigma_{\mathrm{p}} v_{\mathrm{p}}\left(p + p_{\mathrm{t,D}}\right)}, \tag{18}$$

with $\sigma_{\mathrm{n}}$ ($\sigma_{\mathrm{p}}$) the capture-cross section of electrons (holes), $v_{\mathrm{n}}$ ($v_{\mathrm{p}}$) the thermal velocity of electrons (holes), and $n_{\mathrm{t,D}}$ ($p_{\mathrm{t,D}}$) the concentration of electrons (holes) emitted from a donor-like trap [22].

The carrier transport equations

$$\mathbf{J}_{\mathrm{n}} = |e| n \mu_{\mathrm{n}} \mathbf{E}_{\mathrm{n}} + |e| D_{\mathrm{n}} \nabla n \tag{19a}$$

and

$$\mathbf{J}_{\mathrm{p}} = |e| p \mu_{\mathrm{p}} \mathbf{E}_{\mathrm{p}} - |e| D_{\mathrm{p}} \nabla p \tag{19b}$$

are based on the drift-diffusion model [22], with $D_{\mathrm{n}}$ ($D_{\mathrm{p}}$) the diffusion coefficients of electrons

(holes). When space-charge-limited currents dominate charge transport, the second term, i.e., the diffusion current, is negligible compared to the drift current contribution. The current density, as well as the assumed Shockley-Read-Hall-like recombination, are central to the carrier continuity equations [22]:

$$\frac{\partial n}{\partial t} = \frac{1}{|e|}\nabla \cdot \mathbf{J}_\mathrm{n} + (G_\mathrm{n} - R_\mathrm{n}) \tag{20a}$$

and

$$\frac{\partial p}{\partial t} = -\frac{1}{|e|}\nabla \cdot \mathbf{J}_\mathrm{p} + \left(G_\mathrm{p} - R_\mathrm{p}\right), \tag{20b}$$

with $G_\mathrm{n}$ ($G_\mathrm{p}$) and $R_\mathrm{n}$ ($R_\mathrm{p}$) the generation and recombination rates for electrons (holes). These equations comprise the system of coupled, nonlinear partial differential equations which are discretized over the mesh using the finite-element method. The equations are then solved iteratively using the Newton method. The boundaries of the simulation use Neumann boundary conditions, i.e., the current is set to zero [22]. In our simulations, rather than assuming all dopants to be ionized, which is a good assumption at room temperature, the incompletely-ionized dopant concentrations are calculated and used when determining the charge densities.

Finally, mobility models are used in order to more accurately capture the measured *I*–*V* characteristics, in particular with increasing bias which leads to a field-dependent mobility. We use the Lombardi CVT model [58] for the transverse-field dependent component of the mobility, the Klaassen model for the bulk mobility [59, 60], which is in particular relevant for impurity scattering effects in a large concentration range [22], and the Caughey–Thomas saturation-velocity dependent mobility model for the parallel-field component, as given by Equation (8).

### 4.6. Assessment of alternative charge transport mechanisms

Other bulk charge transport mechanisms have been considered in our analysis. Contact-limited mechanisms are excluded due to the Ohmic contacts; the Ohmic *I–V* characteristics of the devices prior to dry etching also support this line of reasoning.

#### 4.6.1. The quasi-exponential regime

Plotting the *I–V* characteristics on a semi-logarithmic scale is a common approach when characterizing diodes as it highlights both the rectifying nature of the device, as well as the exponential nature of the diffusion current. The TLM structures show that the *I–V* characteristic between two terminals is non-rectifying (Figure S10), which is to be expected for a $p^{+}pp^{+}$ structure, assuming that there are no other contacts that break this symmetry. Considering the energy barrier that has formed across the device is thermally activated with $|e|\phi_{\mathrm{B}} \sim E_{\mathrm{a}}$ and that the charge density, and consequently, the current is related exponentially to the applied voltage:

$$I \propto \exp\left(-\frac{E_{\mathrm{a}}}{k_{\mathrm{B}}T} + \frac{|e|V}{\eta k_{\mathrm{B}}T}\right). \tag{21}$$

Extracting the slope in the exponential-like regime ($T = 300$ K) yields ideally factors $\eta > 2$, which indicates that a pure diffusion or recombination current, or a combination of the two currents, does not adequately describe transport in this regime [34].

We also investigate the physical origin of this regime using TCAD simulations (Figure S11). The hole concentration $p$, assuming that the Fermi level is several $k_{\mathrm{B}}T$ away from the valence band edge and Boltzmann statistics can be used, is given by Equation (6). Equating this to the net doping $|N_{\mathrm{A}}^{-} - N_{\mathrm{D}}^{+}|$ yields the difference of the Fermi level and the valence band edge $E_{\mathrm{Fp}} - E_{\mathrm{v}}$ at which the space-charge effect occurs, and thus the onset of SCLCs is expected, given that the drift current contribution dominates the overall current density:

$$E_{\mathrm{Fp}} - E_{\mathrm{v}}\left(p = N_{\mathrm{A}}^{-} - N_{\mathrm{D}}^{+}\right) = k_{\mathrm{B}}T \ln\left(\frac{N_{\mathrm{v}}}{|N_{\mathrm{A}}^{-} - N_{\mathrm{D}}^{+}|}\right). \tag{22}$$

At 300 K, given $N_\mathrm{v} = 1.021 \times 10^{19}$ cm$^{-3}$ [22], $N_\mathrm{D}^+ = 10^{15}$ cm$^{-3}$ and $N_\mathrm{A}^- \approx 5.7 \times 10^{14}$ cm$^{-3}$ (both $N_\mathrm{A}^-$ and $E_\mathrm{Fp} - E_\mathrm{v}$ are extracted in the center of the device at a depth where the hole concentration is largest), this corresponds to $E_\mathrm{Fp} - E_\mathrm{v} = 261$ meV (see black dashed line in Figure S11). This condition should hold across the entire device to speak of the current being space-charge-limited.

While the strong modulation of the electrostatic potential and thus of $E_\mathrm{Fp} - E_\mathrm{v}(V)$ resembles an exponential dependence in the *I–V* characteristics, the nonlinear nature of $E_\mathrm{Fp} - E_\mathrm{v}$ below $p = \left|N_\mathrm{A}^- - N_\mathrm{D}^+\right|$ rules out exponential behavior corresponding to a pure diffusion current. Figure S11 additionally highlights that the voltage at which the transition to the SCLC regime occurs (based on the condition $p = \left|N_\mathrm{A}^- - N_\mathrm{D}^+\right|$) is indeed well approximated by the crossover voltage $V_\mathrm{cross} = V|_{\max(m)}$ (here equal to ~1.1 V). The depth at which the carrier concentration is maximum thus provides an estimate of the region where most of the current flows; however, small deviations (also as a function of channel length) introduce a systematic error in the extraction of $N_\mathrm{eff}$ in Figure 3c, 4a, S17b, and S18.

#### *4.6.2. Variable-range hopping*

The nonlinear nature of the current-voltage characteristics in the treated samples was originally fully attributed to a two-dimensional VRH mechanism close to the silicon surface due to an extracted $T^{-\frac{1}{3}}$ dependence of the resistance [10]. While this was explained by hopping along dopants, given the nature of the silicon surface, hopping along interface states should also not be excluded. However, further measurements and analysis on treated samples could not conclusively establish any particular temperature dependence that would point toward nearest-neighbor hopping or VRH (Figure S12). Fitting in an identical window as in Ref. [10] shows strong correlations for all temperature dependencies; similar analysis and conclusions can be made with the available data from Ref. [10]. Thus, we can speak of Arrhenius behavior, as well as an apparent transition to a freeze-out regime with decreasing temperature, but hopping mechanisms cannot be proven with certainty.

*4.6.3. Poole-Frenkel emission*

An alternative mechanism such as Poole–Frenkel emission [34, 61] across the interface traps was considered:

$$J_\mathrm{s} \propto \mathcal{E} \exp\left[\frac{-|e|\left(\phi_{\mathrm{B,PF}} - \sqrt{\frac{|e|\mathcal{E}}{\pi\epsilon}}\right)}{k_\mathrm{B}T}\right], \tag{23}$$

where $J_\mathrm{s}$ is the surface current density flowing along the interface traps and $\phi_{\mathrm{B,PF}}$ is the barrier height which corresponds to the depth of the traps' potential well. The *I–V* characteristics of the TLM structure in Figure 3 are re-illustrated in the form of Poole–Frenkel plots (Figure S13a), assuming a uniform electric field $\mathcal{E} = V/L$ across the device. Fits performed for both intermediate fields and in the high-field limit resulted in either physically non-viable dielectric constants ($\epsilon_r < 1$), or permittivities exceeding that of silicon (Figure S13b). Furthermore, given the dimensions of both the devices and test structures, Poole–Frenkel emission is unlikely to be the dominant transport mechanism. This model also assumes the absence of space-charge effects and therefore a uniform electric field, whereas the present analysis indicates that space-charge effects are present in the device.

The strong experimental evidence, supported by TCAD simulations, indicating a bulk conduction mechanism reduces the likelihood of charge transport being dominated by mechanisms based on localized surface states, such as Poole–Frenkel emission or 2D VRH.

### 4.7. Influence of mobility on simulated *I*–*V* characteristics

A deeper understanding of the length dependence of the current density can be obtained from TCAD simulations. Normalizing the *I*–*V* characteristics to different channel lengths (Figure S14) supports the experimental finding (Figure 3b) that, in the SCLC regime, the current density scales as $L^{-1}$. This behavior justifies the inclusion of the Caughey–Thomas mobility model to account for field-dependent mobility.

In contrast, assuming a field-independent mobility leads to a current density scaling of $L^{-2}$ in the SCLC regime (Figure S15), as predicted by Equation (7). In this case, the quadratic voltage dependence of the Mott–Gurney law is recovered (Figure S16).

### 4.8. Crossover voltage extraction

As discussed in Section 2.3, the peak voltage in the $m(V)$ characteristics is used to extract the crossover voltage $V_{\text{cross}}$. The quantity $m(V)$ is obtained by locally estimating the power-law exponent via a sliding-window fit. Specifically, least-squares linear regression is performed in a window of size $M$ in $\log I$–$\log V$ space, which is sequentially shifted across the data. This yields a set of local slopes $m$, each associated with an averaged voltage

$$\frac{1}{M}\sum_{k=0}^{M-1} V_{i+k}\,, \tag{24}$$

where $i$ denotes the index of the measured $I$–$V$ data. For the extraction of $V_{\text{cross}}$ in Figure 3c, 4a, and S17b, we use $M = 2$ to maximize sensitivity to the local power-law behavior. An exception is one structure in Figure 3c, for which $M = 5$ is used to mitigate noise. For all other $m(V)$ curves shown for visualization purposes, $M = 2$ is employed, except in Figure 3g and 3h, where $M = 20$ and $M = 5$ are used, respectively, to further suppress noise.

## 4.9. Analysis on arsenic-implanted TLM structures

Similar to hole-transport devices implanted with boron, the analysis followed in Figure 3 can be extended to arsenic-implanted devices and test structures. Figure S17a shows the *I–V* characteristics of an As TLM structure, from which we can perform a similar analysis, extracting the power-law exponents and plotting $V_{\mathrm{cross}}$ versus $L^2$ (Figure S17b). The power-law exponent as a function of voltage reveals, similar to the boron-implanted devices, a low-bias Ohmic regime, a strongly nonlinear regime, and a transition toward a velocity-saturation-limited SCLC regime. Similar to the boron-implanted devices, $N_{\mathrm{eff}} = 3.5 \times 10^{15}\ \mathrm{cm}^{-3} \sim |N_{\mathrm{A}}^{-} - N_{\mathrm{D}}^{+}|$, close to the resistivity of the boron-doped wafers (5–10 Ω·cm), which corresponds to $N_{\mathrm{A}}^{-}$ in the range $1 \times 10^{15}$ to $3 \times 10^{15}\ \mathrm{cm}^{-3}$. The extracted $N_{\mathrm{eff}}$ lies closer to the median value of $N_{\mathrm{A}}^{-}$ for these devices, in contrast to the difference observed between $N_{\mathrm{eff}}$ and $N_{\mathrm{D}}^{+}$ in the boron-implanted case. We attribute this primarily to differences in $L_{\mathrm{c}}$. Longer contacts promote an earlier onset of injected-carrier-dominated space charge in the semiconductor by enabling increased carrier injection (Section 4.10.1).

## 4.10. Influence of various parameters on space-charge effect

### *4.10.1. Geometry*

We also investigate how the contact length $L_{\mathrm{c}}$, i.e., the length of the metal in the *x*-direction (along the channel), as in Figure 4b, influences the crossover voltage, and thus the electrostatics in the device. Figure S18 shows a clear trend of the crossover voltage decreasing with increasing $L_{\mathrm{c}}$. Due to the higher concentration of carriers injected into the device for the longer contacts, the condition for which $p > |N_{\mathrm{A}}^{-} - N_{\mathrm{D}}^{+}|$ shifts to lower voltages. This therefore also influences the extraction of $N_{\mathrm{eff}}$, which decreases with increasing $L_{\mathrm{c}}$. We thus conclude that the geometry of the contacts is central to how space charge is spread across the device. This also expresses the need for a three-dimensional model in order to adequately describe the electrostatics in RNPUs.

#### *4.10.2. Background doping*

TCAD simulations also allow us to observe how the background doping influences the device nonlinearity. Figure S19a shows that the crossover voltage increases with increasing doping concentration, as does the degree of nonlinearity. Since the net doping increases, the depletion region formed at the *p*-*n* junction shifts upwards, decreasing the equilibrium hole concentration. The injected holes must counteract more dopants across the device such that the voltage at which the SCLC regime is entered naturally shifts to higher voltages. The sub-Ohmic behavior observed for $N_{\mathrm{D}}^{+} = 10^{16}\ \mathrm{cm}^{-3}$ can be explained by impurity and carrier-carrier scattering effects, accounted for by the low-field Klaassen mobility model [22, 59]. When the background doping is *p*-type rather than *n*-type, the nonlinearity is significantly suppressed. Due to the increased equilibrium hole concentration and no depletion region at the *p*-*n* junction to set a low hole concentration, the transition to the velocity-saturation SCLC regime is gradual without any significant rise in current. This is experimentally validated in Figure S19b by fabricating TLM structures identical to those analyzed in Figure 3, but on a *p*-type substrate ($N_{\mathrm{A}}^{-} \sim 2 \times 10^{15}\ \mathrm{cm}^{-3}$). The resulting devices exhibit significantly higher currents compared to those in Figure 3 and show a complete absence of the quasi-exponential transport regime.

#### *4.10.3. Implanted doping profile*

As remarked in Section 2.1, in all TCAD simulations, we utilize the dopant profile based on SIMS measurements, and assume that an effective fraction of dopants to be electrically active.

We assume in TCAD simulations that after RTA, only 20% of dopants are electrically active, and that after RIE, only 5% of the dopants in the channel region are electrically active. Though it is unlikely that the percentage of active dopants is constant versus depth, the analysis here still provides qualitative insight into the necessary assumption of dopant deactivation. Furthermore, based on our analysis in Section 4.1, it is more likely that closer to 100% of dopants are active below the contacts, and that less than 1% of dopants are active in the RIE-exposed channel region. The assumed profile in TCAD thus acts as an effective active-dopant profile, constrained by the measured crossover voltages. We thus analyze the sensitivity of the space-charge effect [34] on the implanted dopant profile, in other words, on $N_{\mathrm{A}}^{-}$, to evaluate the role of dopant deactivation. Note that in the TCAD simulations we assume that $N_{\mathrm{A}} = N_{\mathrm{A}}^{-}$, which does not affect the analysis on the space-charge effect, but only influences the modeling of carrier mobility, which we do not focus on here.

Figure S20a shows that a finite degree of dopant deactivation must be assumed in order to reproduce the experimental observations. The ionized acceptor concentration in the plasma-exposed channel region is crucial (Figure S20b,c), and both $m(V)$ and $V_{\mathrm{cross}}$ are sensitive to the availability of holes, particularly in the channel. In principal, $N_{\mathrm{A}}^{-}$ should be $\sim N_{\mathrm{D}}^{+}$ in the bulk region where the hole concentration is highest, i.e., away from the trap-dense surface where carriers are immobilized. Fundamentally, $p_{\mathrm{eq}}$ must be sufficiently low for the space-charge effect to occur; both surface states and dopant deactivation, which are linked to the RIE process, play an important role in this regard.

#### *4.10.4. Substrate potential*

TCAD simulations support the observation in Ref. [10] that nonlinearity can be introduced to devices with $m \sim 1$ by applying a positive substrate potential on the *n*-type substrate of boron

RNPUs (Figure S21a). TCAD simulations show that the positive substrate potential reduces the free hole concentration in the bulk, increasing the depletion width at the *p*-*n* junction between the implanted region and the *n*-doped substrate (Figure S21b). Due to the lower potentials on the input and output electrodes, holes move toward the contacts, decreasing the equilibrium hole concentration $p_{\mathrm{eq}}$ in the bulk (Figure S21c). The reduction in $p_{\mathrm{eq}}$ accounts for the observed shift of $V_{\mathrm{cross}}$ to higher voltages with increasing $V_{\mathrm{sub}}$, as shown in Figure S21a. Notably, $\frac{\partial p}{\partial \ell}$ increases with increasing $V_{\mathrm{sub}}$, which signifies a stronger contribution of the diffusion current to the current density. This supports our discussions and analysis in Sections 2.3 and 2.4 that point towards a large role of diffusion current in the quasi-exponential regime. The substrate potential can thus be seen as effectively enhancing the internal electric field counteracting carrier injection, leading to a strongly nonuniform electric field profile which facilitates steep carrier concentration gradients, manifesting as a pronounced nonlinear dependence of current density on voltage.

### 4.11. Statement on the use of AIGC tools

We have solely used ChatGPT for language and code improvement.

**Acknowledgments**

We thank A. Rop, A. Delke, F. Reenders, M. H. Siekman, M. Schremb and D.H. Wielens for technical support, F. Taglietti and M. Fanciulli for stimulating discussions. We also acknowledge T.T.H. Vu, S. Roy, and A. Attariabad for guidance on the use of measurement equipment, and M. Wiegerinck for contributions to TCAD simulations. We acknowledge tascon GmbH for performing the SIMS measurements and analysis shown in Figure S2.

### Conflicts of Interest

The authors declare no conflicts of interest.

### Data availability statement

All data and code underlying this study are openly available [62].

## References

1. J. D. Kendall, and S. Kumar. The Building Blocks Of A Brain-Inspired Computer. *Applied Physics Reviews*, *7*, no. 1, 011305, 2020. https://doi.org/10.1063/1.5129306.

2. M. A. Zidan, J. P. Strachan, and W. D. Lu. The Future Of Electronics Based On Memristive Systems. *Nature Electronics*, *1*, no. 1, 22–29, 2018. https://doi.org/10.1038/s41928-017-0006-8.

3. M. Xu, X. Chen, Y. Guo, et al. Reconfigurable Neuromorphic Computing: Materials, Devices, And Integration. *Advanced Materials*, *35*, no. 51, 2301063, 2023. https://doi.org/10.1002/adma.202301063.

4. M.-K. Song, J.-H. Kang, X. Zhang, et al. Recent Advances And Future Prospects For Memristive Materials, Devices, And Systems. *ACS Nano*, *17*, no. 13, 11994–12039, 2023. https://doi.org/10.1021/acsnano.3c03505.

5. S. Choi, T. Moon, G. Wang, and J. J. Yang. Filament-Free Memristors For Computing. *Nano Convergence*, *10*, no. 1, 58, 2023. https://doi.org/10.1186/s40580-023-00407-0.

6. T. Günkel, E. Miranda, L. Balcells, N. Mestres, A. Palau, and J. Suñé. Trap-Controlled Conduction And Metal–Insulator Transition In Superconducting Cuprate Memristors. *ACS Applied Electronic Materials*, *8*, no. 3, 1099–1107, 2026. https://doi.org/10.1021/acsaelm.5c02017.

7. C. Pan, C.-Y. Wang, S.-J. Liang, et al. Reconfigurable Logic And Neuromorphic Circuits Based On Electrically Tunable Two-Dimensional Homojunctions. *Nature Electronics*, *3*, no. 7, 383–390, 2020. https://doi.org/10.1038/s41928-020-0433-9.

8. T. Mikolajick, G. Galderisi, S. Rai, et al. Reconfigurable Field Effect Transistors: A Technology Enablers Perspective. *Solid-State Electronics*, *194*, 108381, 2022. https://doi.org/10.1016/j.sse.2022.108381.

9. D. Guo, M. Jia, Y. Wang, et al. Neuromorphic Silicon-Based Capacitive-Tunneling Junction. *Advanced Materials*, 2416643, 2025. https://doi.org/10.1002/adma.202416643.

10. T. Chen, J. van Gelder, B. van de Ven, et al. Classification With A Disordered Dopant-Atom Network In Silicon. *Nature*, *577*, no. 7790, 341–345, 2020. https://doi.org/10.1038/s41586-019-1901-0.

11. H.-C. Ruiz Euler, M. N. Boon, J. T. Wildeboer, et al. A Deep-Learning Approach To Realizing Functionality In Nanoelectronic Devices. *Nature Nanotechnology*, *15*, no. 12, 992–998, 2020. https://doi.org/10.1038/s41565-020-00779-y.

12. H.-C. Ruiz-Euler, U. Alegre-Ibarra, B. van de Ven, H. Broersma, P. A. Bobbert, and W. G. van der Wiel. Dopant Network Processing Units: Towards Efficient Neural Network Emulators With High-Capacity Nanoelectronic Nodes. *Neuromorphic Computing and Engineering*, *1*, no. 2, 024002, 2021. https://doi.org/10.1088/2634-4386/ac1a7f.

13. M. Zolfagharinejad, J. Büchel, L. Cassola, et al. Analogue Speech Recognition Based On Physical Computing. *Nature*, *645*, no. 8082, 886–892, 2025. https://doi.org/10.1038/s41586-025-09501-1.

14. M. Escudero, M. Zolfagharinejad, S. van den Belt, N. Alachiotis, and W. G. van der Wiel. Physical Analog Kolmogorov-Arnold Networks Based On Reconfigurable Nonlinear-Processing Units. 2026. https://doi.org/10.48550/ARXIV.2602.07518.

15. F. Taglietti, A. Pulici, M. Roxburgh, et al. Learning Nonlinear Heterogeneity In Physical Kolmogorov-Arnold Networks. 2026. https://doi.org/10.48550/arXiv.2601.15340.

16. H. Tertilt, J. Bakker, M. Becker, et al. Hopping-Transport Mechanism For Reconfigurable Logic In Disordered Dopant Networks. *Physical Review Applied*, *17*, no. 6, 064025, 2022. https://doi.org/10.1103/PhysRevApplied.17.064025.

17. H. Tertilt, J. Mensing, M. Becker, W. G. van der Wiel, P. A. Bobbert, and A. Heuer. Critical Nonlinear Aspects Of Hopping Transport For Reconfigurable Logic In Disordered Dopant Networks. *Physical Review Applied*, *22*, no. 2, 024063, 2024. https://doi.org/10.1103/PhysRevApplied.22.024063.

18. T. Chen, J. van Gelder, B. van de Ven, et al. Author Correction: Classification With A Disordered Dopant Atom Network In Silicon. *Nature*, *639*, no. 8056, E22–E22, 2025. https://doi.org/10.1038/s41586-025-08803-8.

19. R. K. Ray, and H. Y. Fan. Impurity Conduction In Silicon. *Physical Review*, *121*, no. 3, 768–779, 1961. https://doi.org/10.1103/PhysRev.121.768.

20. W. N. Shafarman, D. W. Koon, and T. G. Castner. Dc Conductivity Of Arsenic-Doped Silicon Near The Metal-Insulator Transition. *Physical Review B*, *40*, no. 2, 1216–1231, 1989. https://doi.org/10.1103/PhysRevB.40.1216.

21. J. Zhang, W. Cui, M. Juda, et al. Hopping Conduction In Partially Compensated Doped Silicon. *Physical Review B*, *48*, no. 4, 2312–2319, 1993. https://doi.org/10.1103/PhysRevB.48.2312.

22. Victory Device (Version 1.24.0.R). Silvaco, Inc. Santa Clara, CA, USA. 2024. Available at: https://silvaco.com/.

23. J. D. Plummer, M. Deal, and P. B. Griffin. Silicon VLSI Technology: Fundamentals, Practice And Modeling. 2000.

24. T. Pfeil, J. Jordan, T. Tetzlaff, et al. Effect Of Heterogeneity On Decorrelation Mechanisms In Spiking Neural Networks: A Neuromorphic-Hardware Study. *Physical Review X*, *6*, no. 2, 021023, 2016. https://doi.org/10.1103/PhysRevX.6.021023.

25. L. Ribar, and R. Sepulchre. Neuromorphic Control: Designing Multiscale Mixed-Feedback Systems. *IEEE Control Systems Magazine*, *41*, no. 6, 34–63, 2021. https://doi.org/10.1109/MCS.2021.3107560.

26. M. Dragoman, and D. Dragoman. Negative Differential Resistance In Novel Nanoscale Devices. *Solid-State Electronics*, *197*, 108464, 2022. https://doi.org/10.1016/j.sse.2022.108464.

27. H. Lu, S. Xie, W. Zhang, et al. Negative Differential Resistance In Memristive Systems: Historical Evolution, Mechanisms And Neuromorphic Applications Of Niobium Oxide Devices. *Nanoscale*, *17*, no. 36, 20606–20642, 2025. https://doi.org/10.1039/D5NR02491A.

28. J. Kareem. Manuscript In Preparation. 2026.

29. C. R. Helms, and E. H. Poindexter. The Silicon-Silicon Dioxide System: Its Microstructure And Imperfections. *Reports on Progress in Physics*, *57*, no. 8, 791–852, 1994. https://doi.org/10.1088/0034-4885/57/8/002.

30. P. M. Lenahan, T. D. Mishima, J. Jumper, T. N. Fogarty, and R. T. Wilkins. Direct Experimental Evidence For Atomic Scale Structural Changes Involved In The Interface-Trap Transformation Process. *IEEE Transactions on Nuclear Science*, *48*, no. 6, 2131–2135, 2001. https://doi.org/10.1109/23.983184.

31. S. T. Chang, J. K. Wu, and S. A. Lyon. Amphoteric Defects At The Si-SiO2 Interface. *Applied Physics Letters*, *48*, no. 10, 662–664, 1986. https://doi.org/10.1063/1.96736.

32. H. Hasegawa, T. Sato, S. Kasai, B. Adamowicz, and T. Hashizume. Dynamics And Control Of Recombination Process At Semiconductor Surfaces, Interfaces And Nano-Structures. *Solar Energy*, *80*, no. 6, 629–644, 2006. https://doi.org/10.1016/j.solener.2005.10.014.

33. J. B. Varley, J. R. Weber, A. Janotti, and C. G. van de Walle. Dangling Bonds, The Charge Neutrality Level, And Band Alignment In Semiconductors. *Journal of Applied Physics*, *135*, no. 7, 075703, 2024. https://doi.org/10.1063/5.0190043.

34. S. M. Sze, and K. K. Ng. Physics Of Semiconductor Devices. 2007.

35. D. K. Schroder. Semiconductor Material And Device Characterization. 2005. https://doi.org/10.1002/0471749095.

36. G. T. Wright. Mechanisms Of Space-Charge-Limited Current In Solids. *Solid-State Electronics*, *2*, nos. 2–3, 165–189, 1961. https://doi.org/10.1016/0038-1101(61)90034-X.

37. A. A. Grinberg, and S. Luryi. Space-Charge-Limited Current And Capacitance In Double-Junction Diodes. *Journal of Applied Physics*, *61*, no. 3, 1181–1189, 1987. https://doi.org/10.1063/1.338165.

38. J. A. Geurst. Theory Of Space-Charge-Limited Currents In Thin Semiconductor Layers. *Physica Status Solidi (b)*, *15*, no. 1, 107–118, 1966. https://doi.org/10.1002/pssb.19660150108.

39. A. A. Grinberg, S. Luryi, M. R. Pinto, and N. L. Schryer. Space-Charge-Limited Current In A Film. *IEEE Transactions on Electron Devices*, *36*, no. 6, 1162–1170, 1989. https://doi.org/10.1109/16.24363.

40. M. A. Lampert, and P. Mark. Current Injection In Solids. 1970.

41. A. Rose. Space-Charge-Limited Currents In Solids. *Physical Review*, *97*, no. 6, 1538–1544, 1955. https://doi.org/10.1103/PhysRev.97.1538.

42. D. M. Caughey, and R. E. Thomas. Carrier Mobilities In Silicon Empirically Related To Doping And Field. *Proceedings of the IEEE*, *55*, no. 12, 2192–2193, 1967. https://doi.org/10.1109/PROC.1967.6123.

43. K. W. Lee, and Y. S. Ang. Injection-Limited And Space-Charge-Limited Conduction In Wide Bandgap Semiconductors With Velocity Saturation Effect. 2023. https://doi.org/10.48550/arXiv.2308.00955.

44. G. S. Oehrlein. Dry Etching Damage Of Silicon: A Review. *Materials Science and Engineering B*, *4*, nos. 1–4, 441–450, 1989. https://doi.org/10.1016/0921-5107(89)90284-5.

45. J. C. Mikkelsen, and I.-W. Wu. Severe Loss Of Dopant Activity Due To CHF3+CO2 Reactive Ion Etch Damage. *Applied Physics Letters*, *49*, no. 2, 103–105, 1986. https://doi.org/10.1063/1.97399.

46. J. M. Heddleson, M. W. Horn, S. J. Fonash, and D. C. Nguyen. Effects Of Dry Etching On The Electrical Properties Of Silicon. *Journal of Vacuum Science & Technology B: Microelectronics Processing and Phenomena*, *6*, no. 1, 280–283, 1988. https://doi.org/10.1116/1.584024.

47. P. Mark, and W. Helfrich. Space-Charge-Limited Currents In Organic Crystals. *Journal of Applied Physics*, *33*, no. 1, 205–215, 1962. https://doi.org/10.1063/1.1728487.

48. M. Nikolka, K. Broch, J. Armitage, et al. High-Mobility, Trap-Free Charge Transport In Conjugated Polymer Diodes. *Nature Communications*, *10*, no. 1, 2122, 2019. https://doi.org/10.1038/s41467-019-10188-y.

49. M. Sajedi Alvar, P. W. M. Blom, and G.-J. A. H. Wetzelaer. Space-Charge-Limited Electron And Hole Currents In Hybrid Organic-Inorganic Perovskites. *Nature Communications*, *11*, no. 1, 4023, 2020. https://doi.org/10.1038/s41467-020-17868-0.

50. V. M. Le Corre, E. A. Duijnstee, O. El Tambouli, et al. Revealing Charge Carrier Mobility And Defect Densities In Metal Halide Perovskites Via Space-Charge-Limited Current Measurements. *ACS Energy Letters*, *6*, no. 3, 1087–1094, 2021. https://doi.org/10.1021/acsenergylett.0c02599.

51. P. Zhang, Y. S. Ang, A. L. Garner, Á. Valfells, J. W. Luginsland, and L. K. Ang. Space–Charge Limited Current In Nanodiodes: Ballistic, Collisional, And Dynamical Effects. *Journal of Applied Physics*, *129*, no. 10, 100902, 2021. https://doi.org/10.1063/5.0042355.

52. E. Landi, A. Armigliato, S. Solmi, R. Kögler, and E. Wieser. Electrical Activation Of Boron-Implanted Silicon During Rapid Thermal Annealing. *Applied Physics A: Solids and Surface*, *47*, no. 4, 359–366, 1988. https://doi.org/10.1007/BF00615499.

53. T. E. F. M. Standaert, C. Hedlund, E. A. Joseph, G. S. Oehrlein, and T. J. Dalton. Role Of Fluorocarbon Film Formation In The Etching Of Silicon, Silicon Dioxide, Silicon Nitride, And Amorphous Hydrogenated Silicon Carbide. *Journal of Vacuum Science & Technology, A: Vacuum, Surfaces, and Films*, *22*, no. 1, 53–60, 2004. https://doi.org/10.1116/1.1626642.

54. A. Efremov, B. J. Lee, and K.-H. Kwon. On Relationships Between Gas-Phase Chemistry And Reactive Ion Etching Kinetics For Silicon-Based Thin Films (SiC, SiO2 And SixNy) In Multi-Component Fluorocarbon Gas Mixtures. *Materials*, *14*, no. 6, 1432, 2021. https://doi.org/10.3390/ma14061432.

55. H. Jansen, H. Gardeniers, M. D. Boer, M. Elwenspoek, and J. Fluitman. A Survey On The Reactive Ion Etching Of Silicon In Microtechnology. *Journal of Micromechanics and Microengineering*, *6*, no. 1, 14–28, 1996. https://doi.org/10.1088/0960-1317/6/1/002.

56. S. K. Bose, C. P. Lawrence, Z. Liu, et al. Evolution Of A Designless Nanoparticle Network Into Reconfigurable Boolean Logic. *Nature Nanotechnology*, *10*, no. 12, 1048–1052, 2015. https://doi.org/10.1038/nnano.2015.207.

57. H. Fukuda, M. Yasuda, T. Iwabuchi, S. Kaneko, T. Ueno, and I. Ohdomari. Process Dependence Of The SiO2/Si(100) Interface Trap Density Of Ultrathin SiO2 Films. *Journal of Applied Physics*, *72*, no. 5, 1906–1911, 1992. https://doi.org/10.1063/1.351665.

58. C. Lombardi, S. Manzini, A. Saporito, and M. Vanzi. A Physically Based Mobility Model For Numerical Simulation Of Nonplanar Devices. *IEEE Transactions on Computer-Aided Design of Integrated Circuits and Systems*, *7*, no. 11, 1164–1171, 1988. https://doi.org/10.1109/43.9186.

59. D. B. M. Klaassen. A Unified Mobility Model For Device Simulation—I. Model Equations And Concentration Dependence. *Solid-State Electronics*, *35*, no. 7, 953–959, 1992. https://doi.org/10.1016/0038-1101(92)90325-7.

60. D. B. M. Klaassen. A Unified Mobility Model For Device Simulation—II. Temperature Dependence Of Carrier Mobility And Lifetime. *Solid-State Electronics*, *35*, no. 7, 961–967, 1992. https://doi.org/10.1016/0038-1101(92)90326-8.

61. R. M. Hill. Poole-Frenkel Conduction In Amorphous Solids. *Philosophical Magazine*, *23*, no. 181, 59–86, 1971. https://doi.org/10.1080/14786437108216365.

62. J. Kareem, L. Cassola, R. J. C. Cool, et al. Code And Data For “Space-Charge-Driven Nonlinear Charge Transport In Silicon Reconfigurable Nonlinear-Processing Units.” 2026. https://doi.org/10.5281/ZENODO.20259201.

**Supporting Information**

for

**Space-Charge-Driven Nonlinear Charge Transport in**

**Silicon Reconfigurable Nonlinear-Processing Units**

J. Kareem[1*], L. Cassola[1], R.J.C. Cool[1,2], J.I. van Slooten[1], R.J.E. Hueting[3], P.A. Bobbert[1,4], W.G. van der Wiel[1,2†]

[1]*NanoElectronics Group, MESA+ Institute, and Center for Brain-Inspired Computing (BRAINS), University of Twente, P.O. Box 217, 7500 AE Enschede, The Netherlands*

[2]*Institute of Physics, University of Münster, 48149 Münster, Germany*

[3]*Power Electronics Group, MESA+ Institute, University of Twente, P.O. Box 217, 7500 AE Enschede, The Netherlands*

[4]*Materials to Optoelectronic Devices and Eindhoven Institute for Renewable Energy Systems, Eindhoven University of Technology, P.O. Box 513, 5600 MB Eindhoven, The Netherlands*

[*]Contact author: jonas.kareem@utwente.nl

[†]Contact author: w.g.vanderwiel@utwente.nl

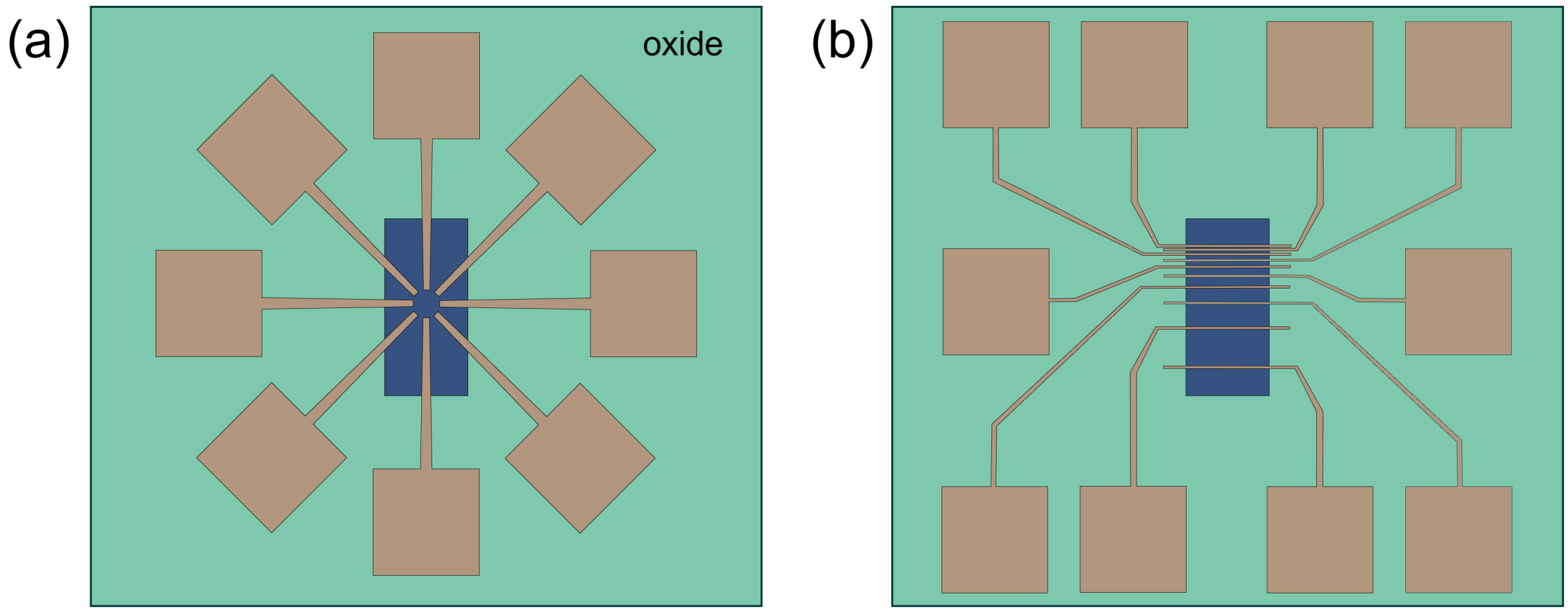

**Figure S1.** Schematic overview of (a) RNPU and (b) TLM designs. Devices are defined on 26 μm × 60 μm implantation window (blue). The 150 μm × 150 μm bonding pads are metalized on 280 nm thick thermally grown silicon oxide (green) covering most of the silicon substrate.

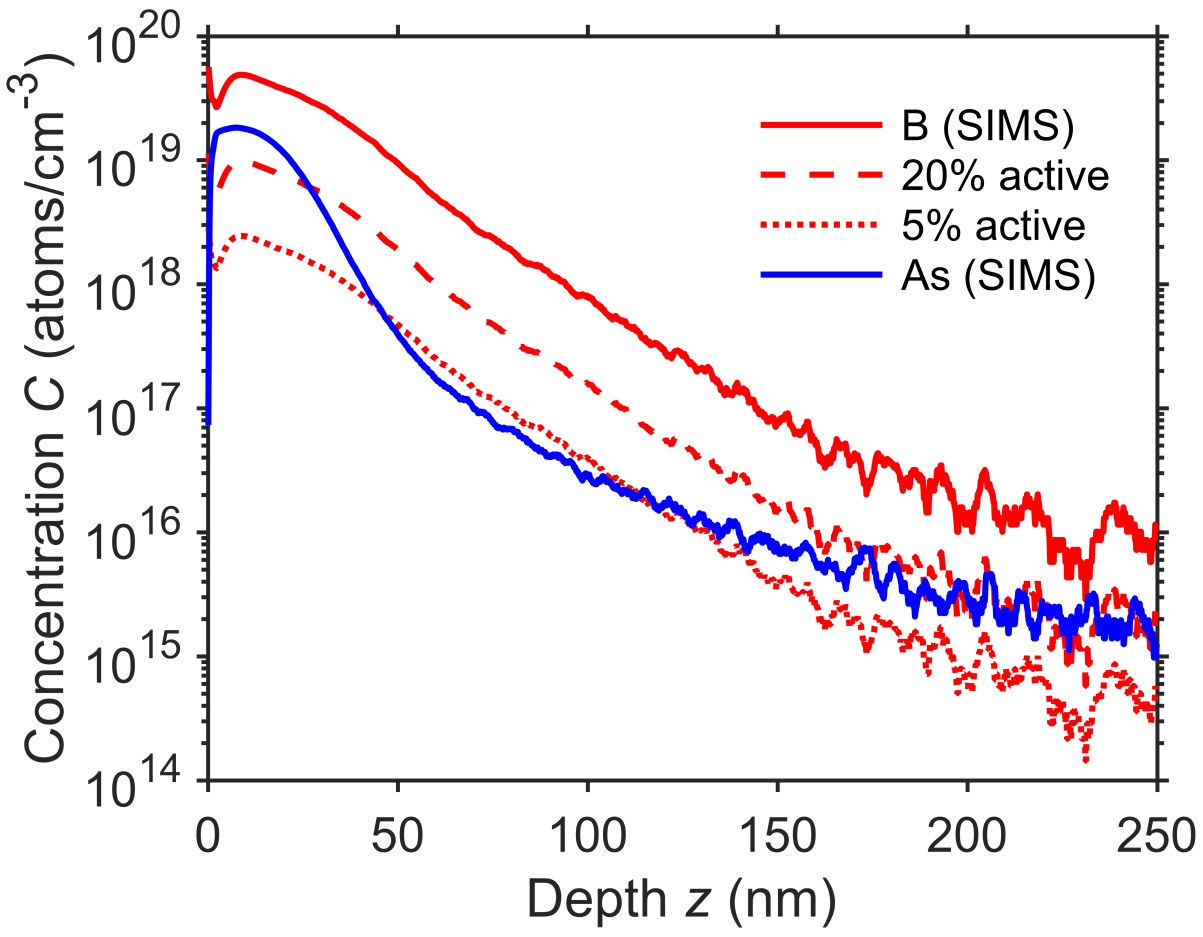

**Figure S2.** Boron (red) and arsenic (blue) implantation concentration *C* profiles as a function of depth *z*, determined via secondary-ion mass spectroscopy (SIMS). The boron dopant profiles used in TCAD simulations for below the contacts (dashed line) and for the channel (dotted line) are also shown, assuming that an effective constant fraction of dopants is inactive. SIMS measurements and analysis performed by tascon GmbH.

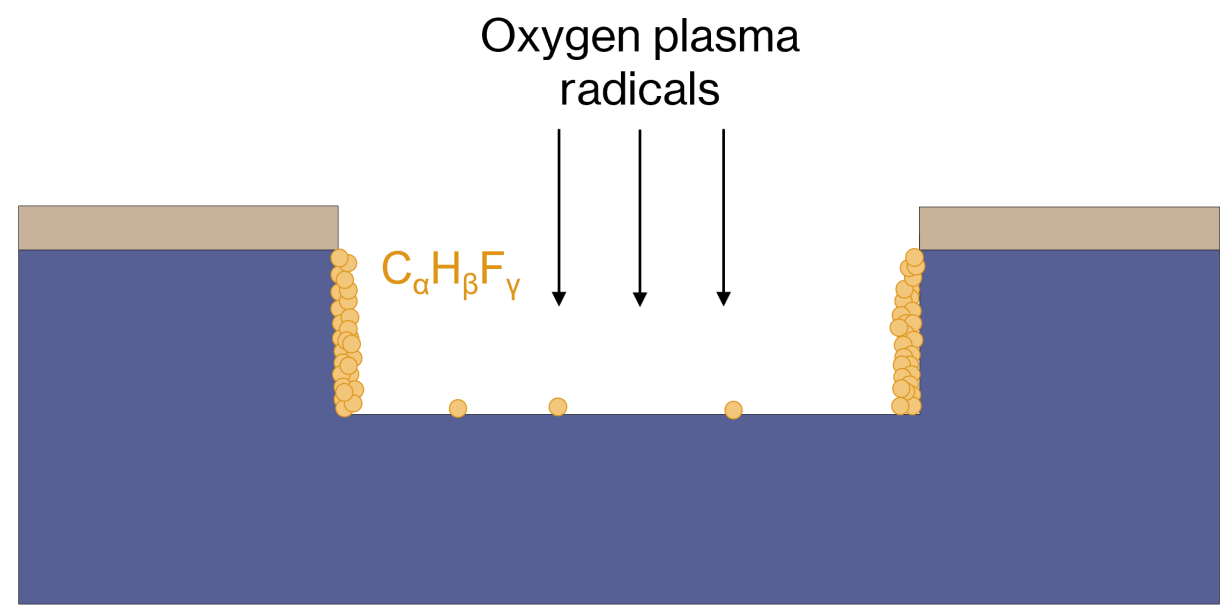

**Figure S3.** Cross-sectional schematic of the reactive-ion etching (RIE) process. $C_\alpha H_\beta F_\gamma$ radicals form a fluorocarbon polymer layer, while oxygen ion bombardment removes this layer from the horizontal (100) surface. Polymer passivation is therefore retained mainly on the sidewalls, protecting them from chemical etching and enabling an anisotropic etch profile.

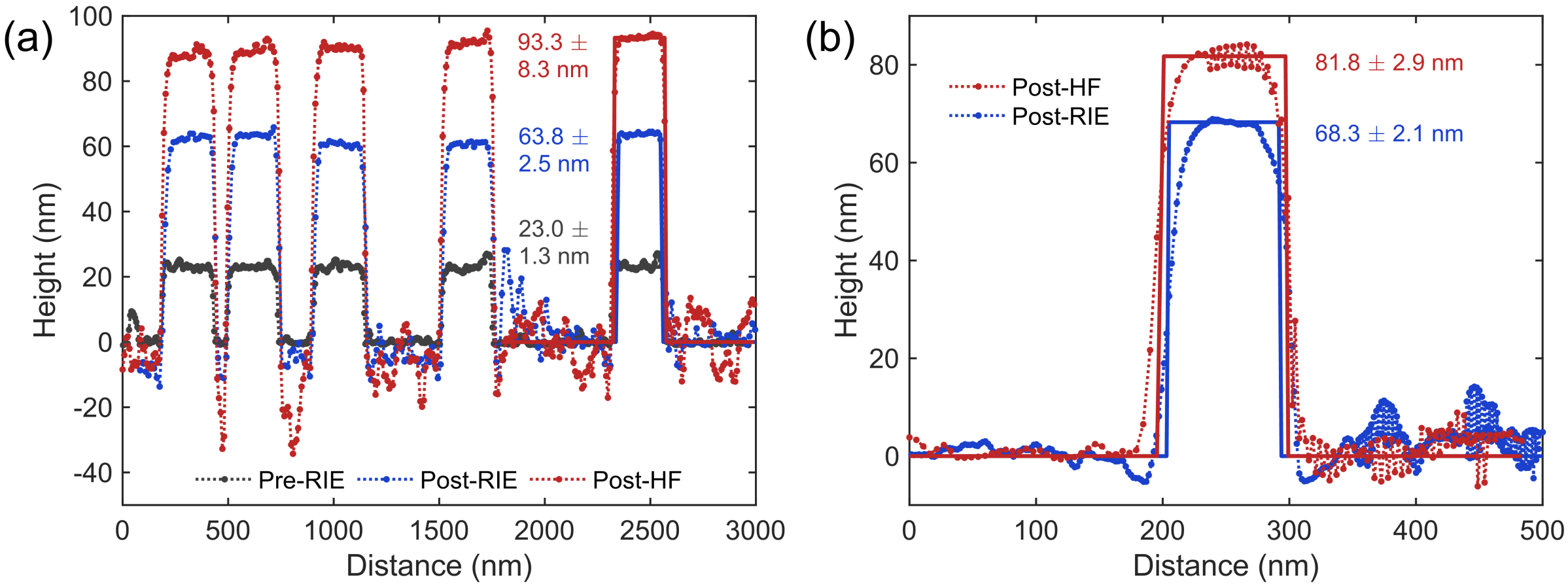

**Figure S4.** (a) Atomic-force-microscope (AFM) cross-section of a TLM structure, showing the measured step heights before RIE, after RIE, and after wet etching in hydrofluoric acid (HF). Based on the difference in step heights before and after HF etching, a 29.5 nm ± 10.8 nm thick fluorocarbon layer is likely deposited on the (100) silicon surface. (b) Height profile across RNPU contacts, where the average step height taken over eight contacts is plotted. The smaller step height difference relative to the TLM structure indicates that the reactive-ion etching is geometry-dependent, with an estimated fluorocarbon layer thickness of 13.5 ± 5.0 nm.

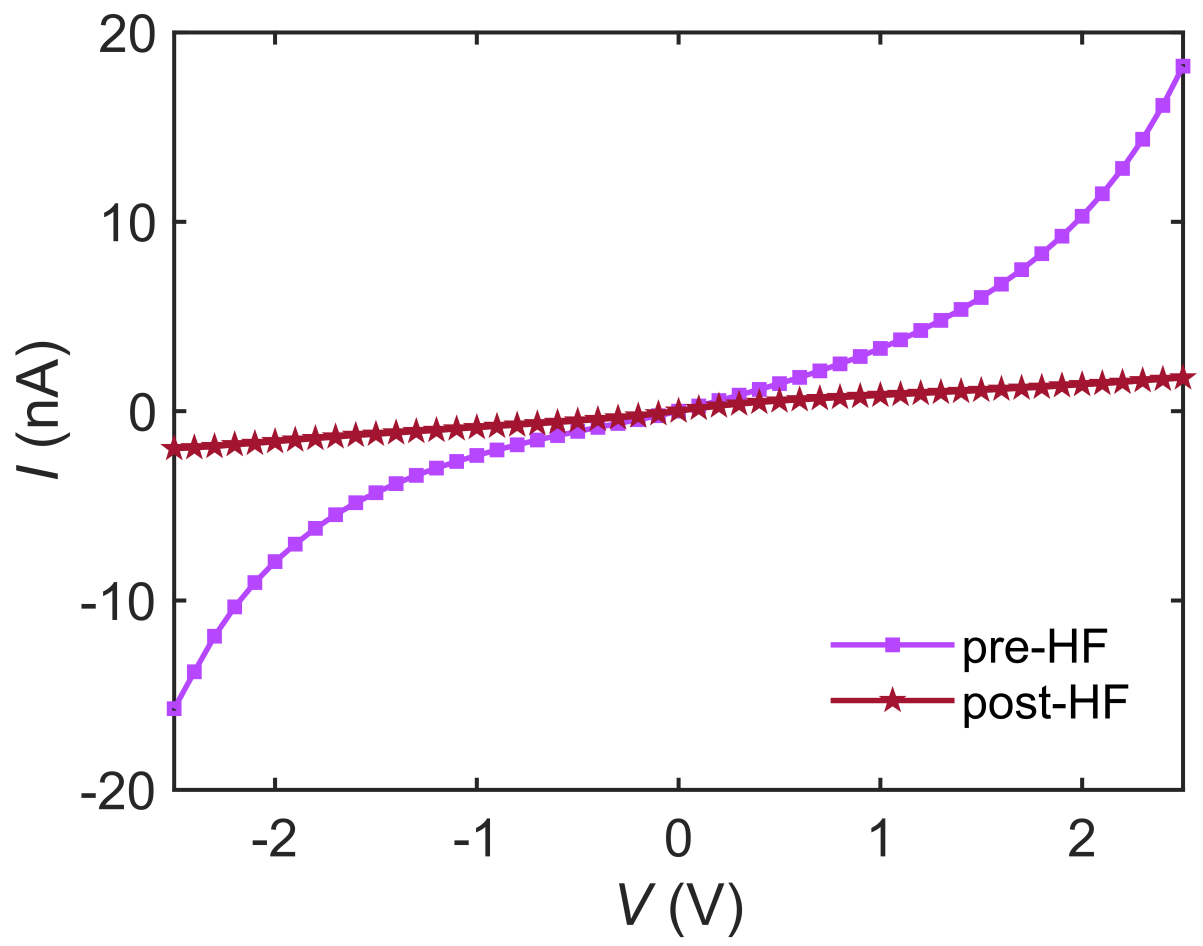

**Figure S5.** Comparison of *I*–*V* characteristics of a boron-implanted RNPU before and after HF wet etch. The bias voltage is swept on one input electrode, the neighboring output electrode is grounded, and the remaining electrodes are left floating.

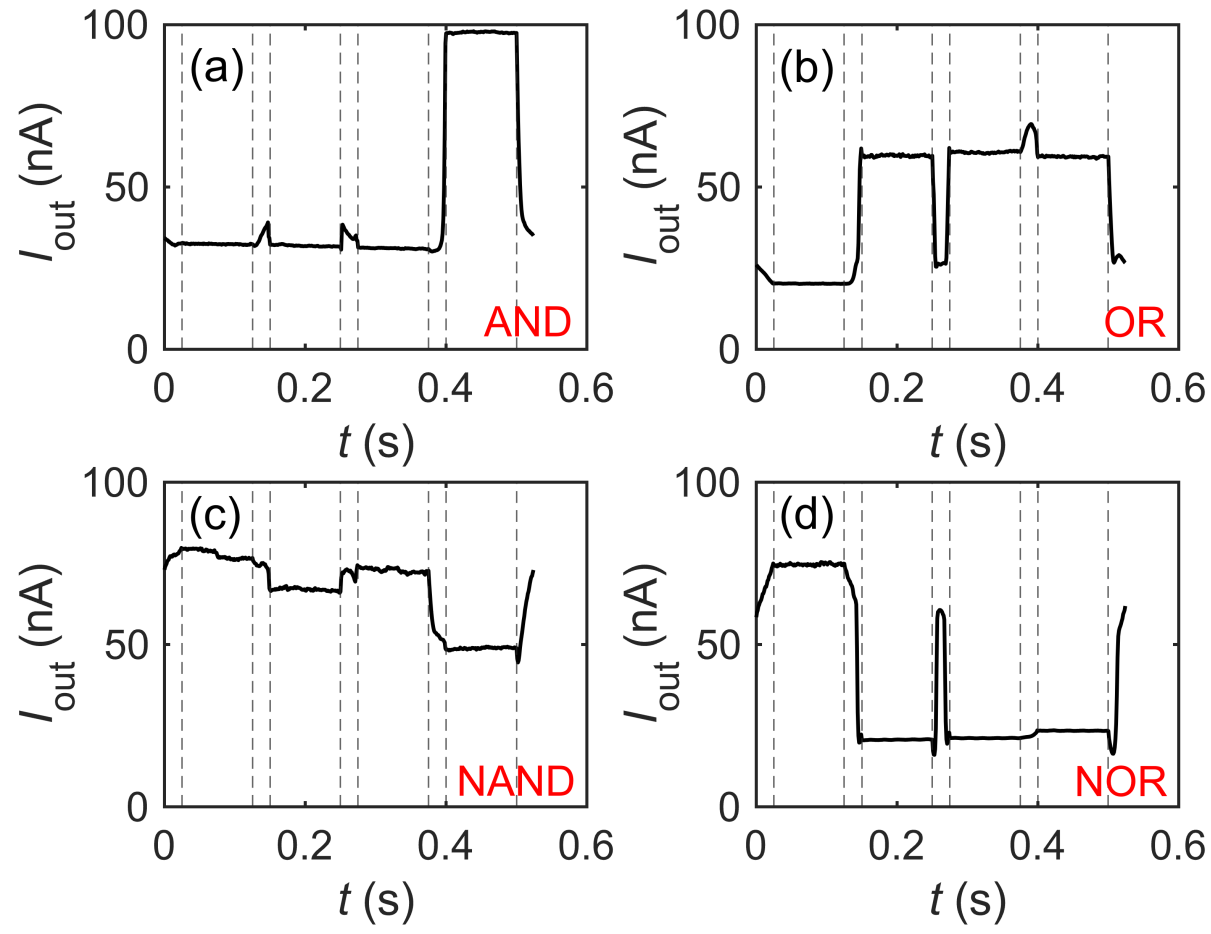

**Figure S6.** Remaining Boolean logic gates solved on same untreated device in Figure 1d. Using Equation (13), the calculated fitness values of the (a) AND, (b) OR, (c) NAND and (d) NOR gates are 9.3, 8.5, 0.4 and 6.2, respectively.

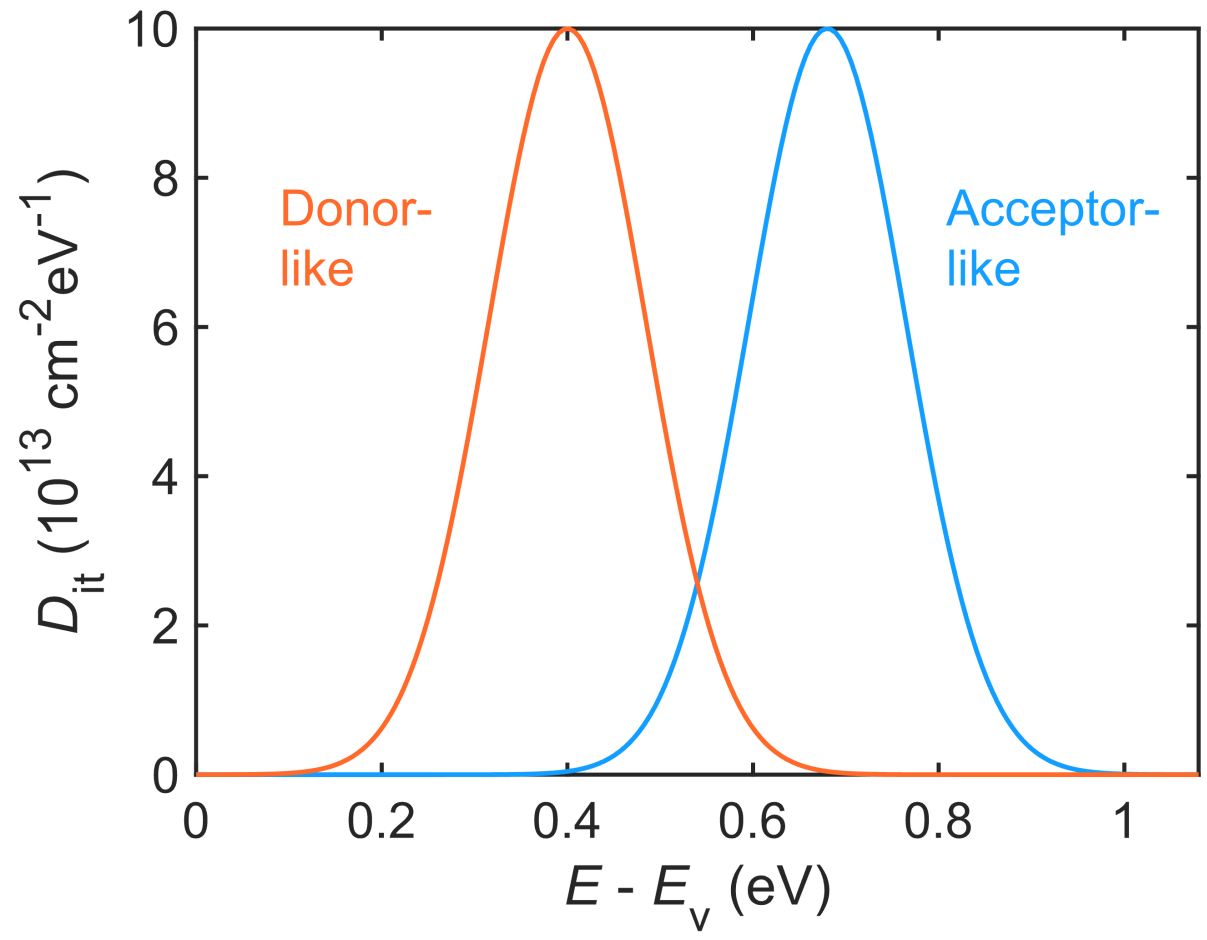

**Figure S7.** Trap distributions used in TCAD [1] at the silicon/silicon oxide interface to emulate $P_b$ centers, approximated here by $P_{b0}$-like states [2, 3].

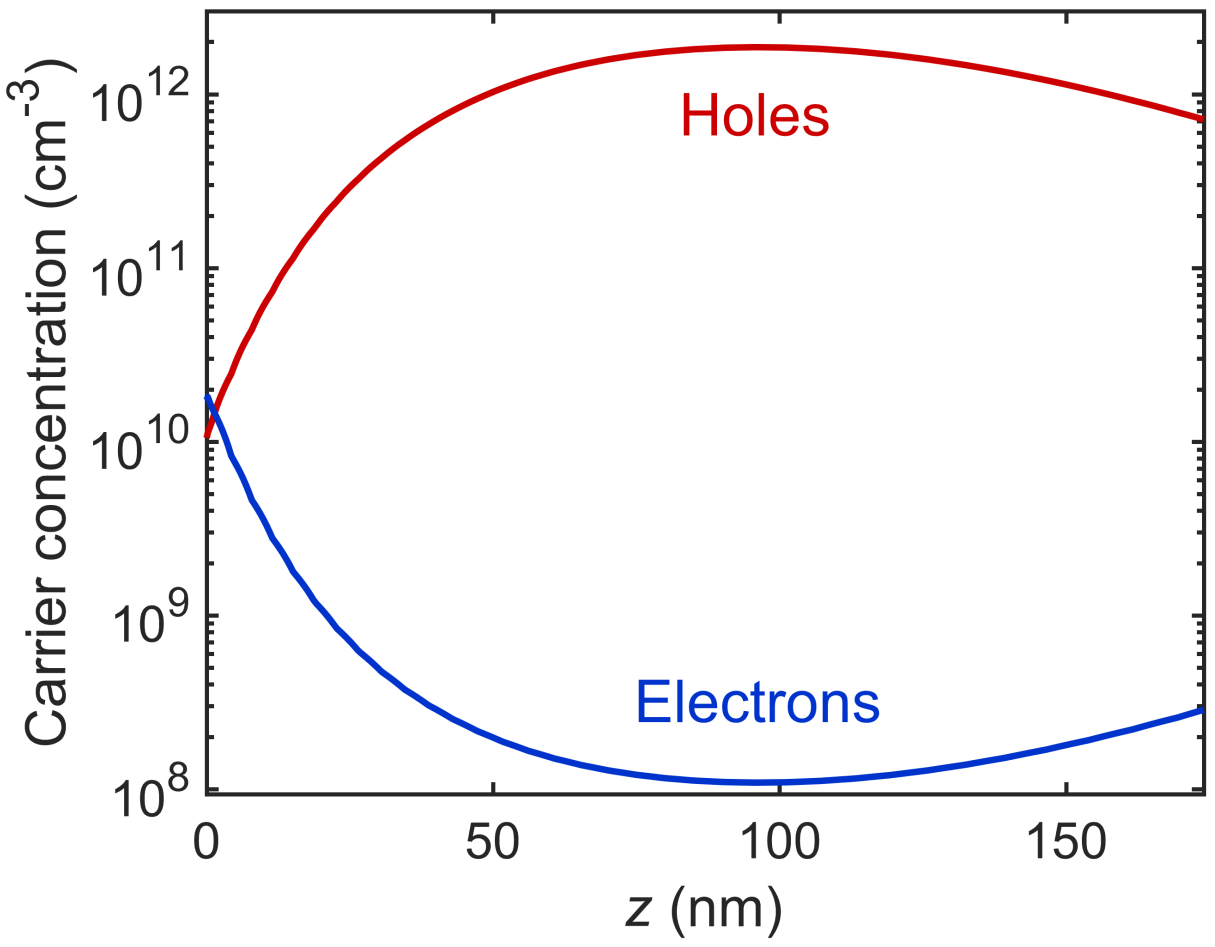

**Figure S8.** Simulated electron and hole concentrations versus depth $z$ in the center of $p$-type TLM structure ($L$ = 300 nm, $L_c$ = 200 nm) at equilibrium. The depth begins at the etched silicon surface with the etch depth equal to 70 nm.

In contrast to the boron-implanted devices, the electron-carrier devices implanted with arsenic show a clear relationship between the etch depth and the extracted activation energy (Figure S9a). By reducing the dopant-induced free carrier concentration via etching, the Fermi level becomes increasingly pinned close to mid-gap. The available free carriers are trapped by the interface states, leaving a lower carrier concentration that may contribute to band transport. As the sheet ionized donor density $N_{\mathrm{D,sheet}}^{+}$ is reduced, the barrier to transport increases in height as a result of the stronger band bending (Figure S9b), and correspondingly the activation energy increases.

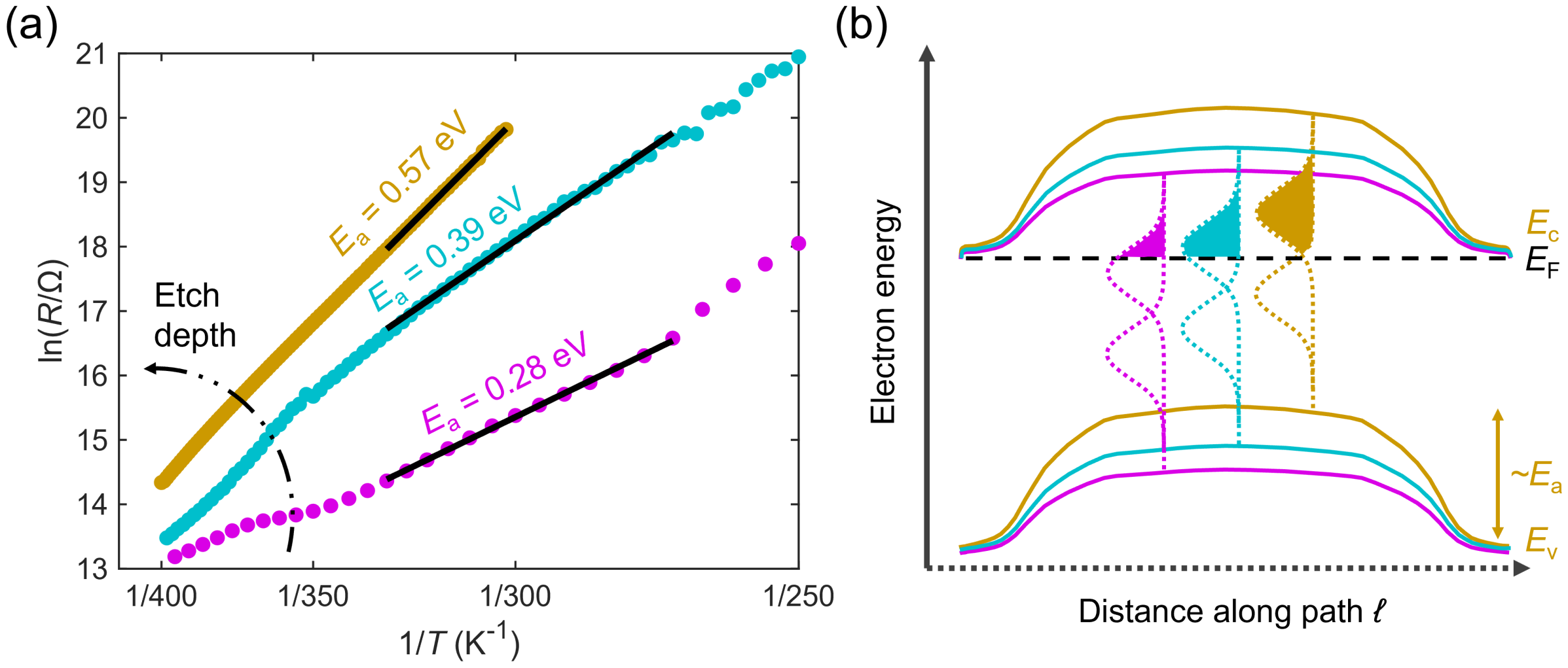

**Figure S9.** (a) Experimental Arrhenius plots of low-bias resistance $R$ (Ω) for arsenic-implanted devices ($I$–$V$ characteristics measured between adjacent contacts) with different etch times and corresponding apparent etch depths (20, 30, and 70 nm). The reported etch depths do not include the fluorocarbon layer (Section 4.1). (b) Equilibrium band diagrams, as in Figure 2b, for the three devices with different etch depths. The distance corresponds to the distance between the contacts, as in Figure 2b.

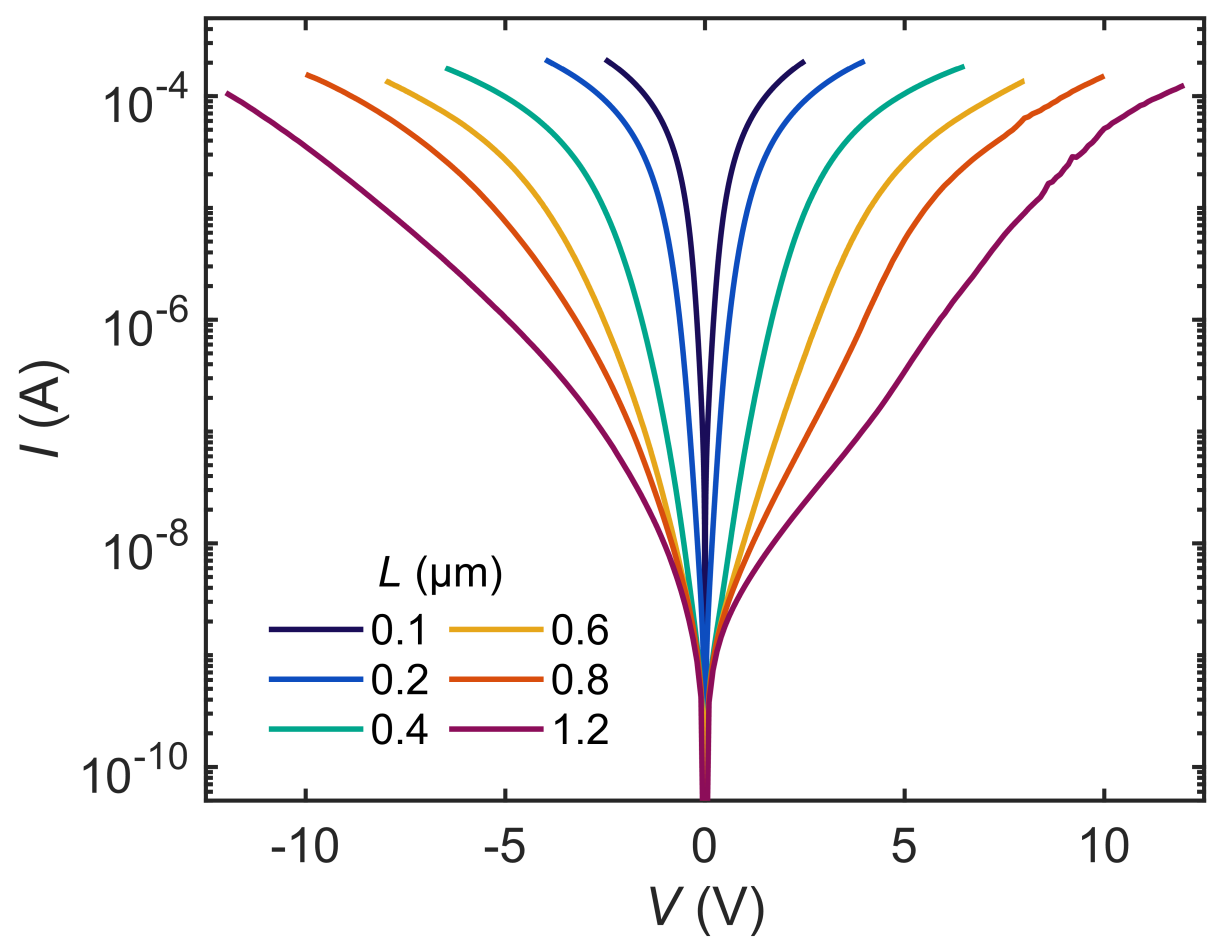

**Figure S10.** Semi-logarithmic plot of $I$–$V$ characteristics in a boron-implanted TLM structure (etch depth = 70 nm) for various $L$.

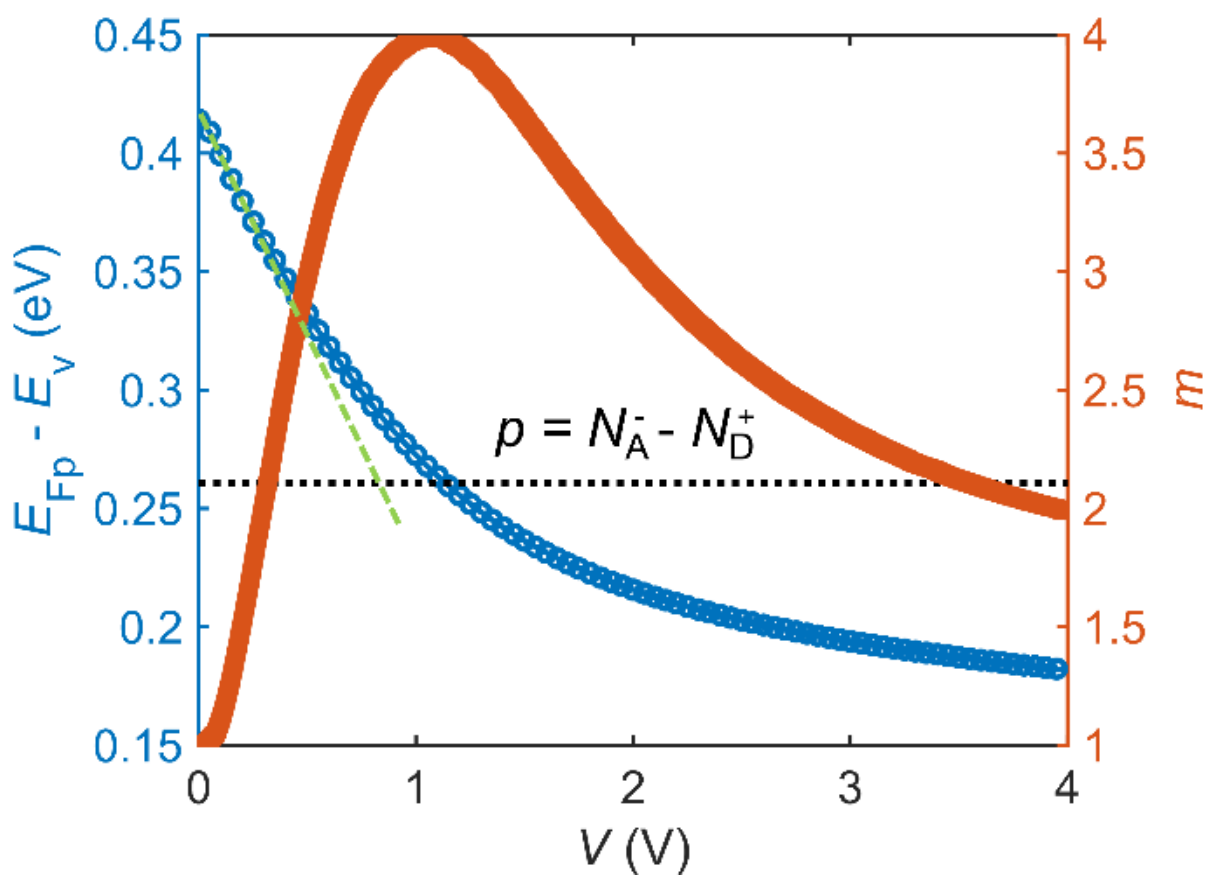

**Figure S11.** TCAD simulation of $E_{\mathrm{Fp}} - E_{\mathrm{v}}$ (left axis) and the power-law exponent (right axis) versus voltage $V$ on linear scale for a $p$-type TLM structure ($L = 300$ nm, $N_{\mathrm{D}}^{+} = 10^{15}$ cm$^{-3}$). The value of $E_{\mathrm{Fp}} - E_{\mathrm{v}}(V)$ at which the free hole concentration $p$ is equal to $\left|N_{\mathrm{A}}^{-} - N_{\mathrm{D}}^{+}\right|$ is represented by the black dotted line. $E_{\mathrm{Fp}} - E_{\mathrm{v}}(V)$ is taken in the center of the device, 150 nm below the etched surface. The crossover voltage $V_{\mathrm{cross}} = 1.1$ V. The light-green dashed line shows that $E_{\mathrm{Fp}} - E_{\mathrm{v}}$ depends nonlinearly on $V$.

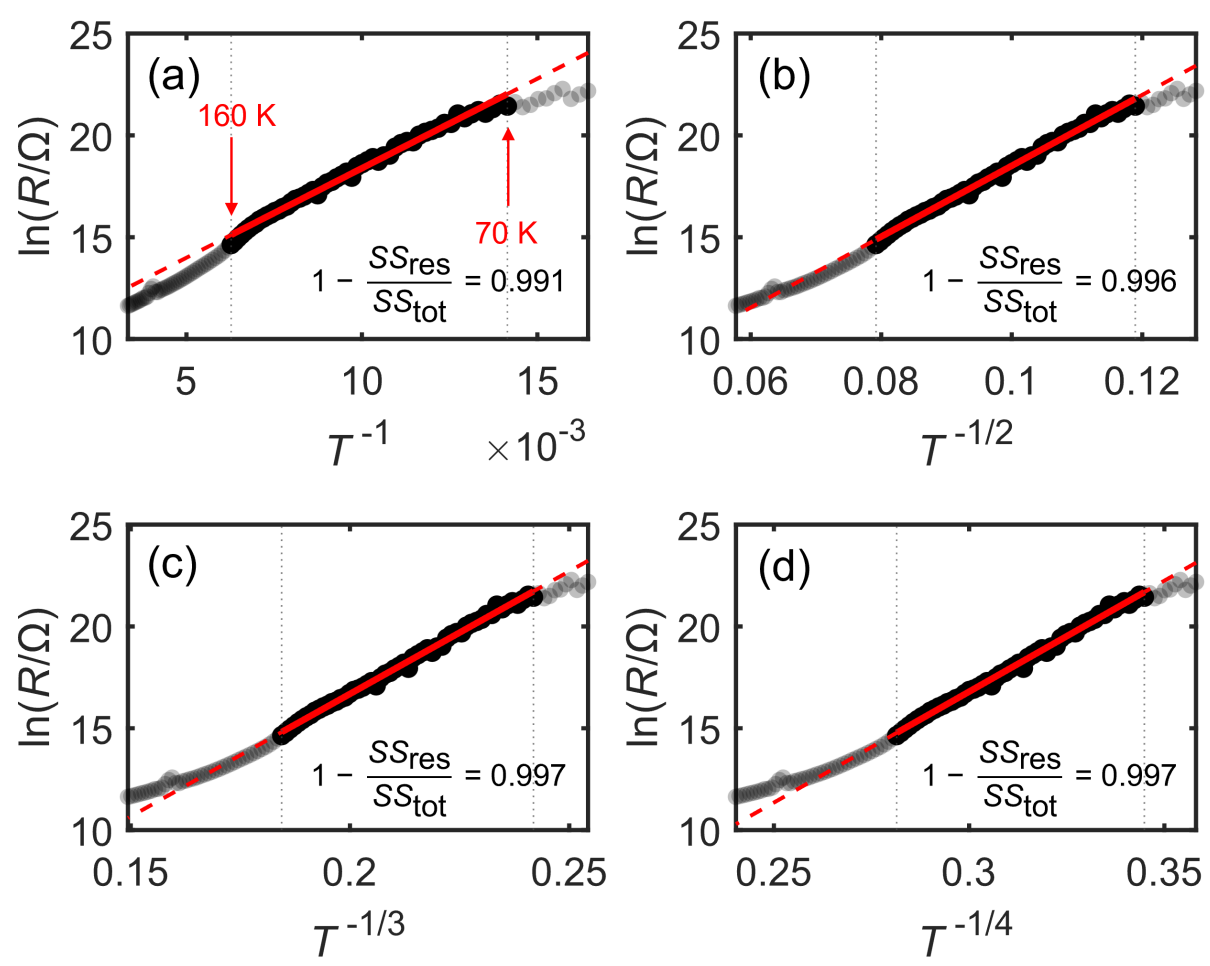

**Figure S12.** Semi-logarithmic plot of low-bias resistance $R$ (Ω) as a function of (a) $T^{-1}$, (b) $T^{-\frac{1}{2}}$, (c) $T^{-\frac{1}{3}}$, and (d) $T^{-\frac{1}{4}}$. The coefficient of determination is given in each plot, with $SS_{\text{res}}$ and $SS_{\text{tot}}$ the residual and total sum of squares, respectively. The sample treatment involved an HF wet etch for 2 minutes, followed by an anneal at 250 ˚C for 5 minutes.

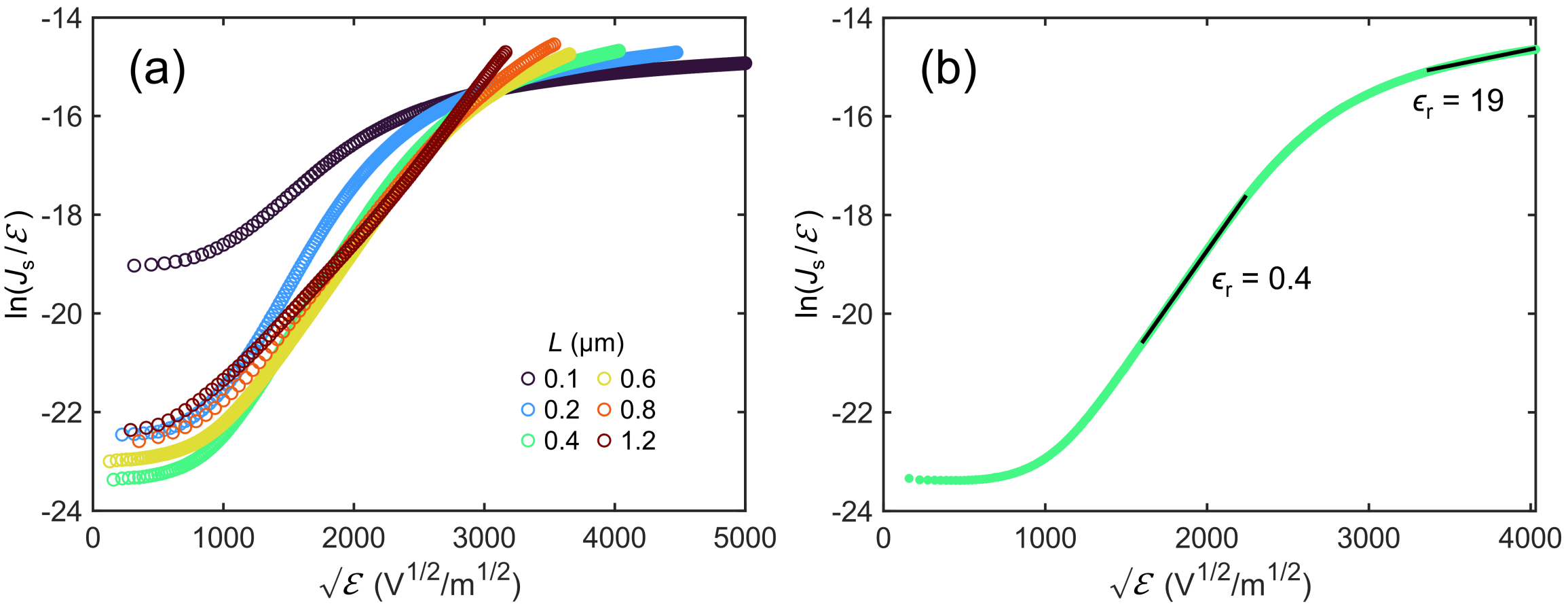

**Figure S13.** (a) Poole–Frenkel plots of untreated boron-implanted TLM structure, plotted for various $L$. The current density is defined as a surface current density $J_s$, normalized to 26 μm, i.e., the width of the TLM structure (Figure S1). The average electric field $\mathcal{E} = V/L$ is assumed to be uniform across the device. (b) Poole–Frenkel plot for a 400 nm channel, including fits in an intermediate-field regime ($\epsilon_r = 0.4$) and in the high-field limit ($\epsilon_r = 19$).

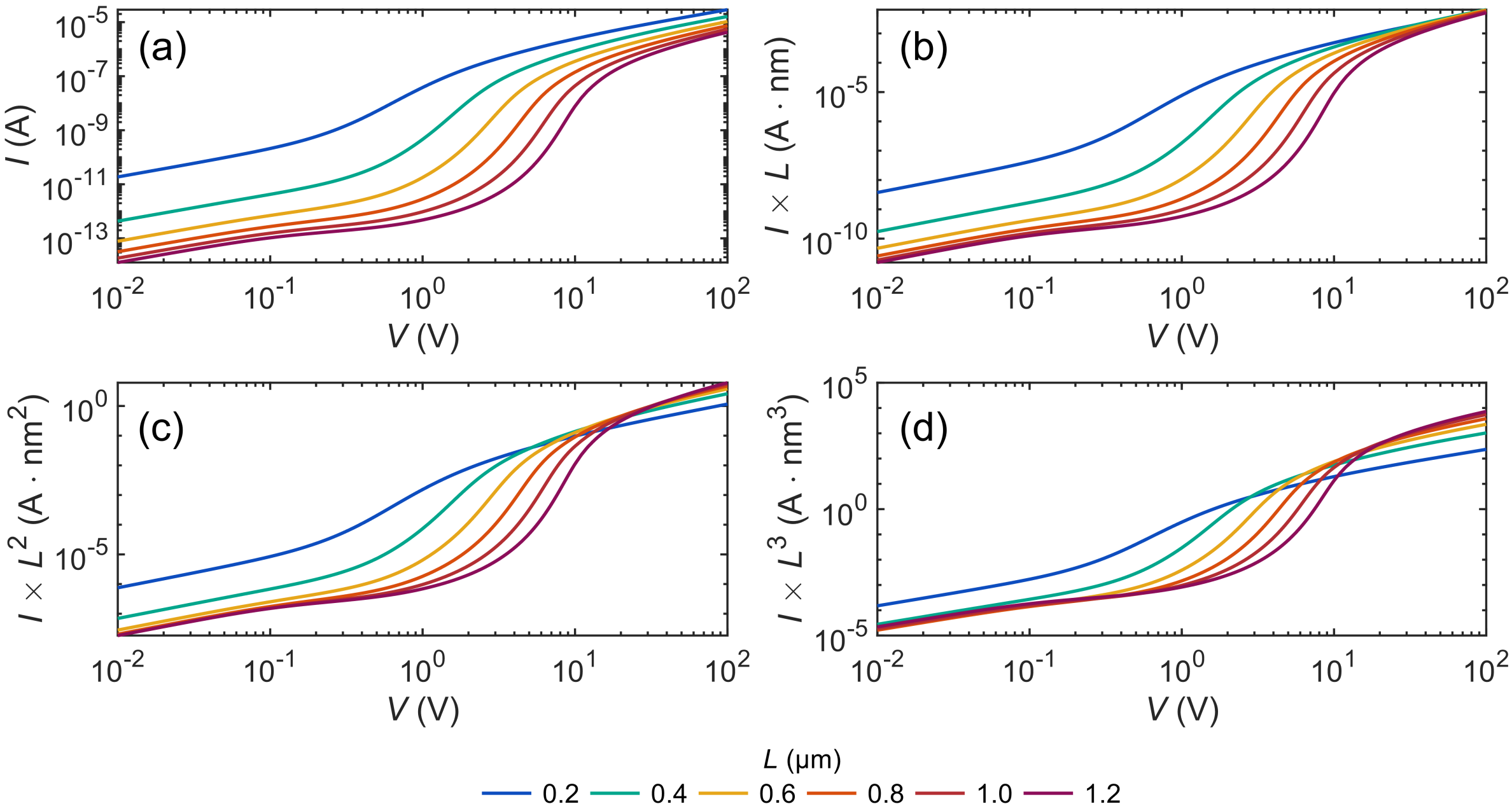

**Figure S14.** (a) $I$–$V$ characteristics of simulated $p$-type TLM structure including all relevant mobility models, normalized to (b) $L$, (c) $L^2$, and (d) $L^3$. The width of the device, which scales the current, is assumed to be 20 nm large. In the high-field regime, the current scales with $L^{-1}$, supporting the theory on SCLC modified by velocity saturation, given by Equation (9), and the experimental results.

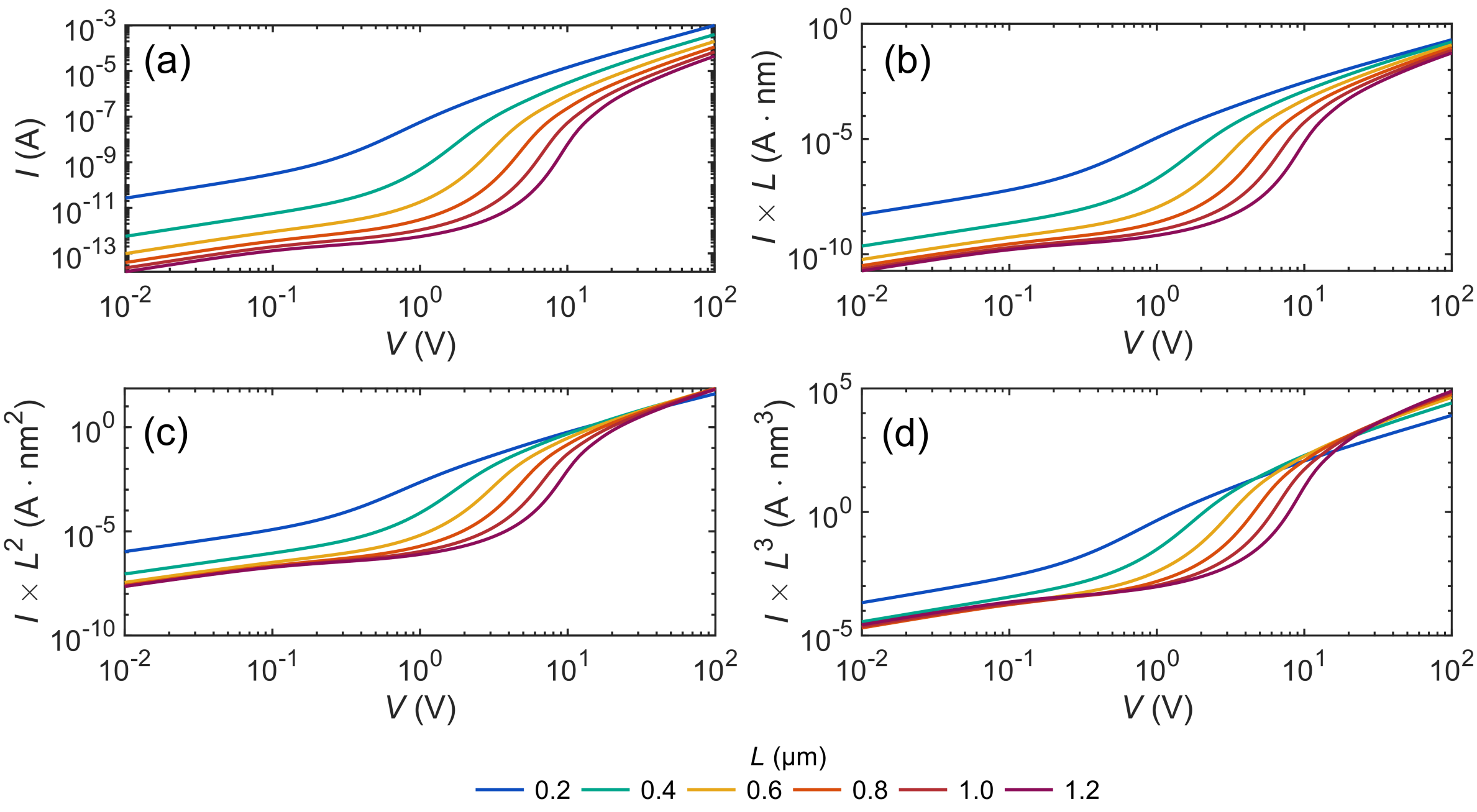

**Figure S15.** (a) $I$–$V$ characteristics of simulated $p$-type TLM structure excluding the Caughey–Thomas mobility model, normalized to (b) $L$, (c) $L^2$, and (d) $L^3$. The width of the device is assumed to be 20 nm large. Since the mobility is independent of the electric field, in the high-field SCLC regime, the current scales with $L^{-2}$, in line with Equation (7).

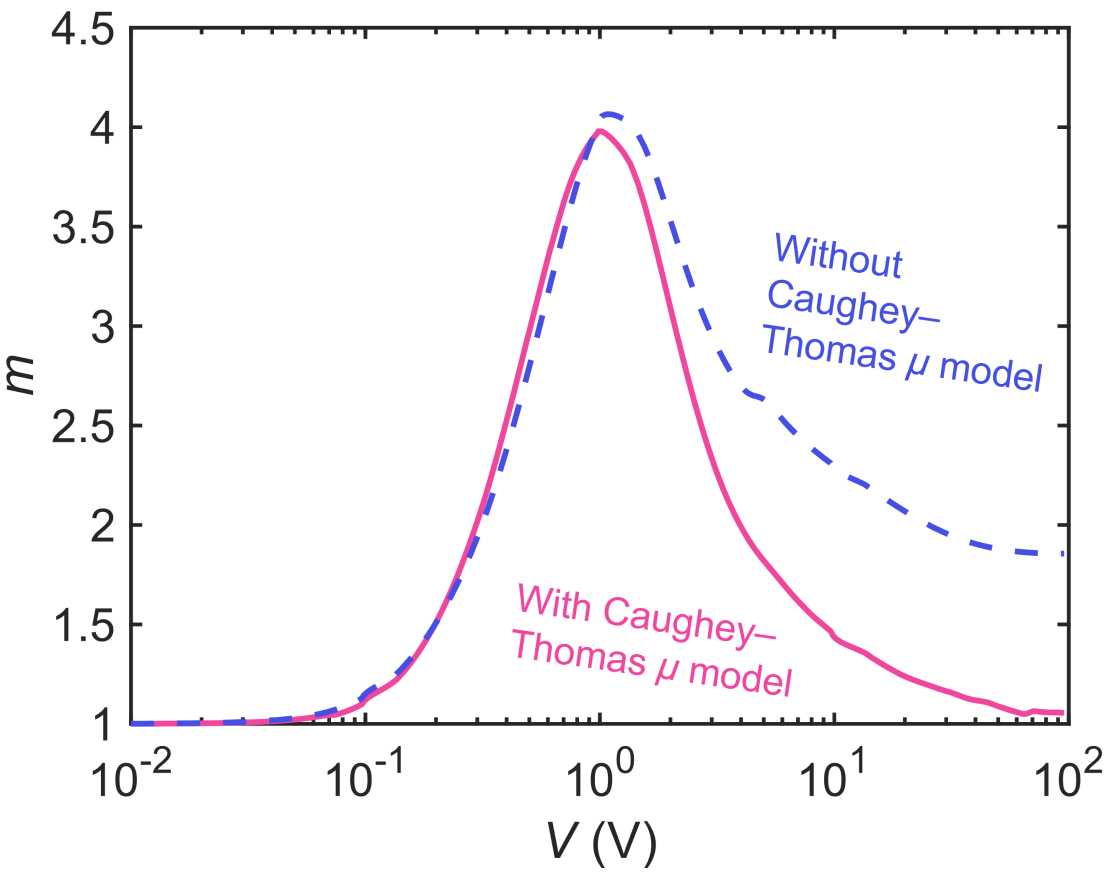

**Figure S16.** Simulated power-law exponent versus voltage in a $p$-type TLM structure ($L$ = 300 nm). The inclusion of mobility models, in particular the Caughey–Thomas mobility model, modifies the voltage dependence such that in the high-field limit, $I \propto V$ instead of $\propto V^2$, supporting Equations (7-9).

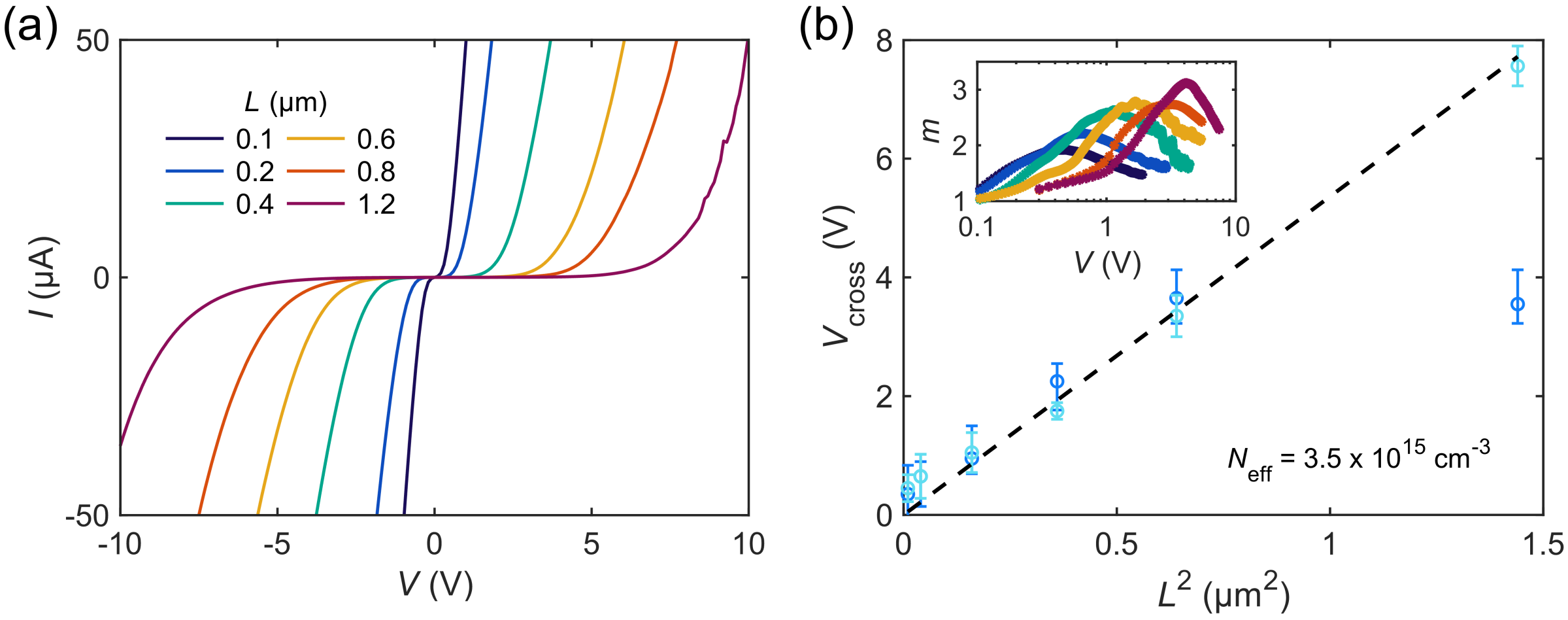

**Figure S17.** (a) TLM $I$–$V$ characteristics on an arsenic-implanted sample with an apparent etch depth equal to 50 nm (etch time = 3 min, $L_c$ = 500 nm). (b) Plot of crossover voltages $V_{\mathrm{cross}}$ versus $L^2$ for two test structures with the geometry described in (a). Based on Equation (11), $N_{\mathrm{eff}} = 3.5 \times 10^{15}\ \mathrm{cm}^{-3}$ is extracted. Inset: Local power-law exponent $m(V)$ for varying channel lengths. An outlier for one 1200 nm channel is excluded from the fit; substrate charging effects may have heavily skewed the crossover voltage prior to and during the measurement.

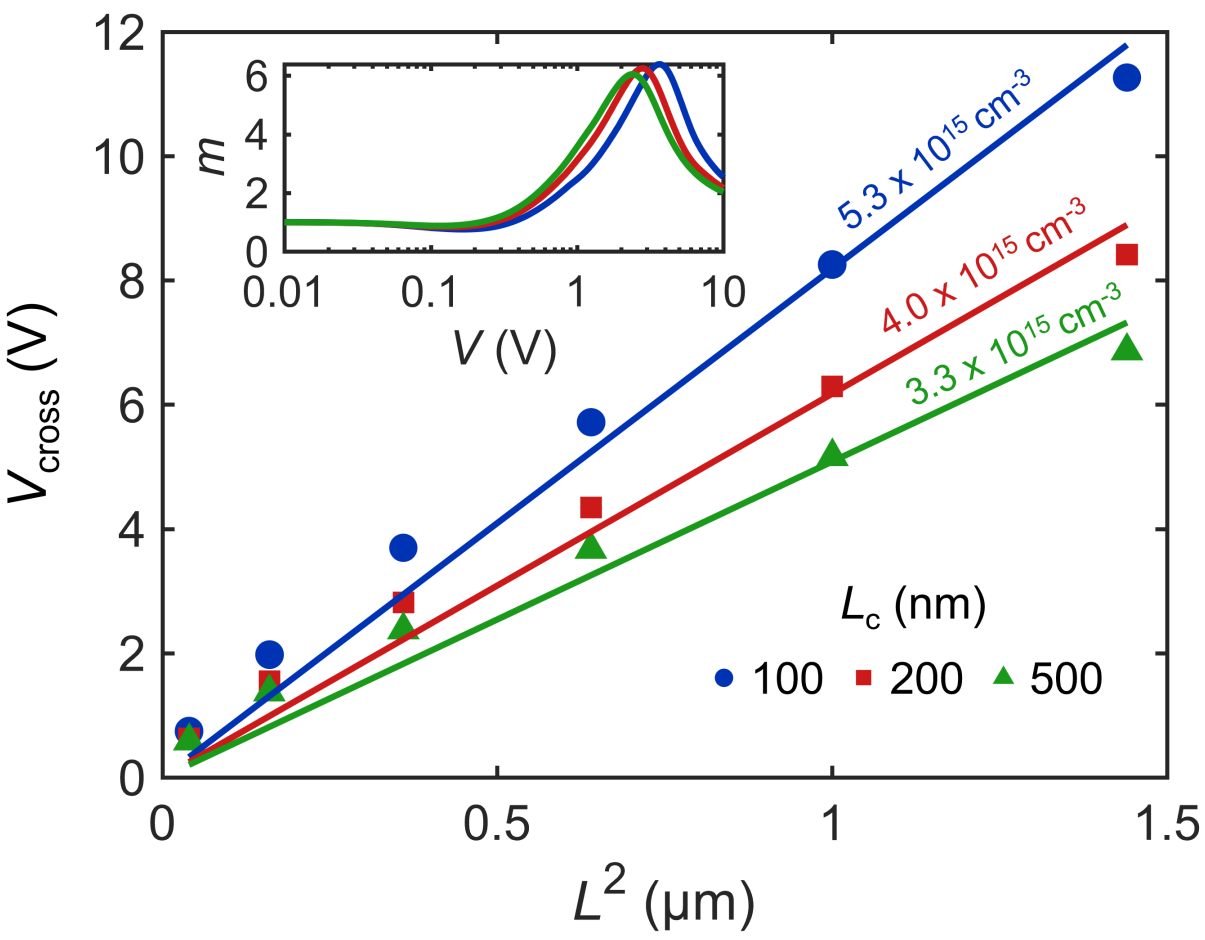

**Figure S18.** Simulated plots of $V_{\mathrm{cross}}$ versus $L^2$ ($N_{\mathrm{D}}^{+} = 10^{15}\ \mathrm{cm}^{-3}$, $N_{\mathrm{tp}} = 10^{12}\ \mathrm{cm}^{-2}\ \mathrm{eV}^{-1}$) for various contact lengths $L_{\mathrm{c}}$ in a $p$-type TLM structure. The extracted $N_{\mathrm{eff}}$ values are indicated alongside the corresponding fitted lines. Inset: Local power-law exponent $m$ versus $V$ for three contact lengths ($L$ = 600 nm).

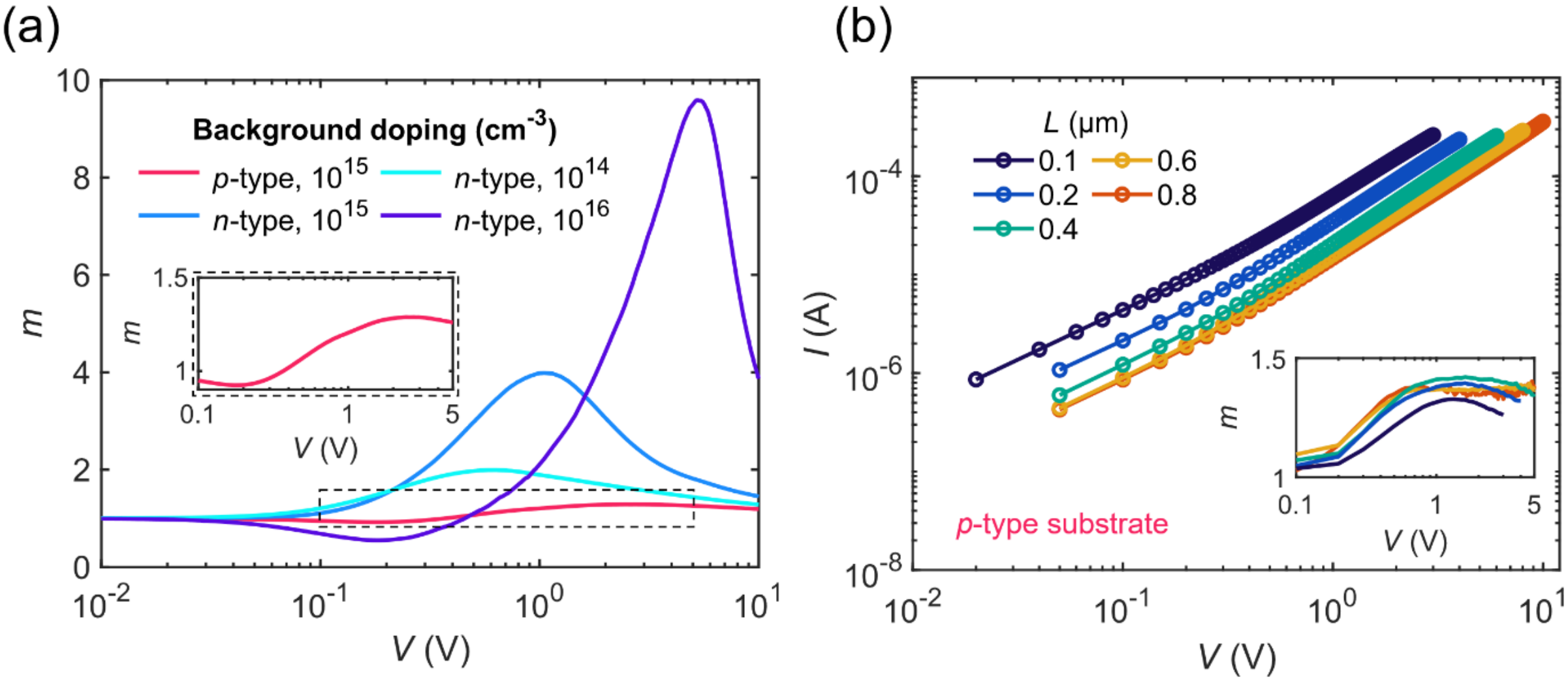

**Figure S19.** (a) Simulated local power-law exponent versus voltage for various background doping concentrations in a *p*-type TLM structure ($L$ = 300 nm, $L_c$ = 200 nm, $N_{tp}$ = $10^{12}$ cm$^{-2}$ eV$^{-1}$). Replacing the *n*-type background doping with *p*-type doping strongly suppresses the nonlinear response. The dashed box highlights the suppressed nonlinearity for a *p*-type background doping concentration of $10^{15}$ cm$^{-3}$. (b) Experimental validation of the simulated behavior in (a). Log–log plot of $I$–$V$ characteristics for a TLM structure ($L_c$ = 200 nm) fabricated identically to the devices in Figure 3, but using a *p*-type substrate. Inset: Local power-law exponent $m$ versus $V$ for various $L$.

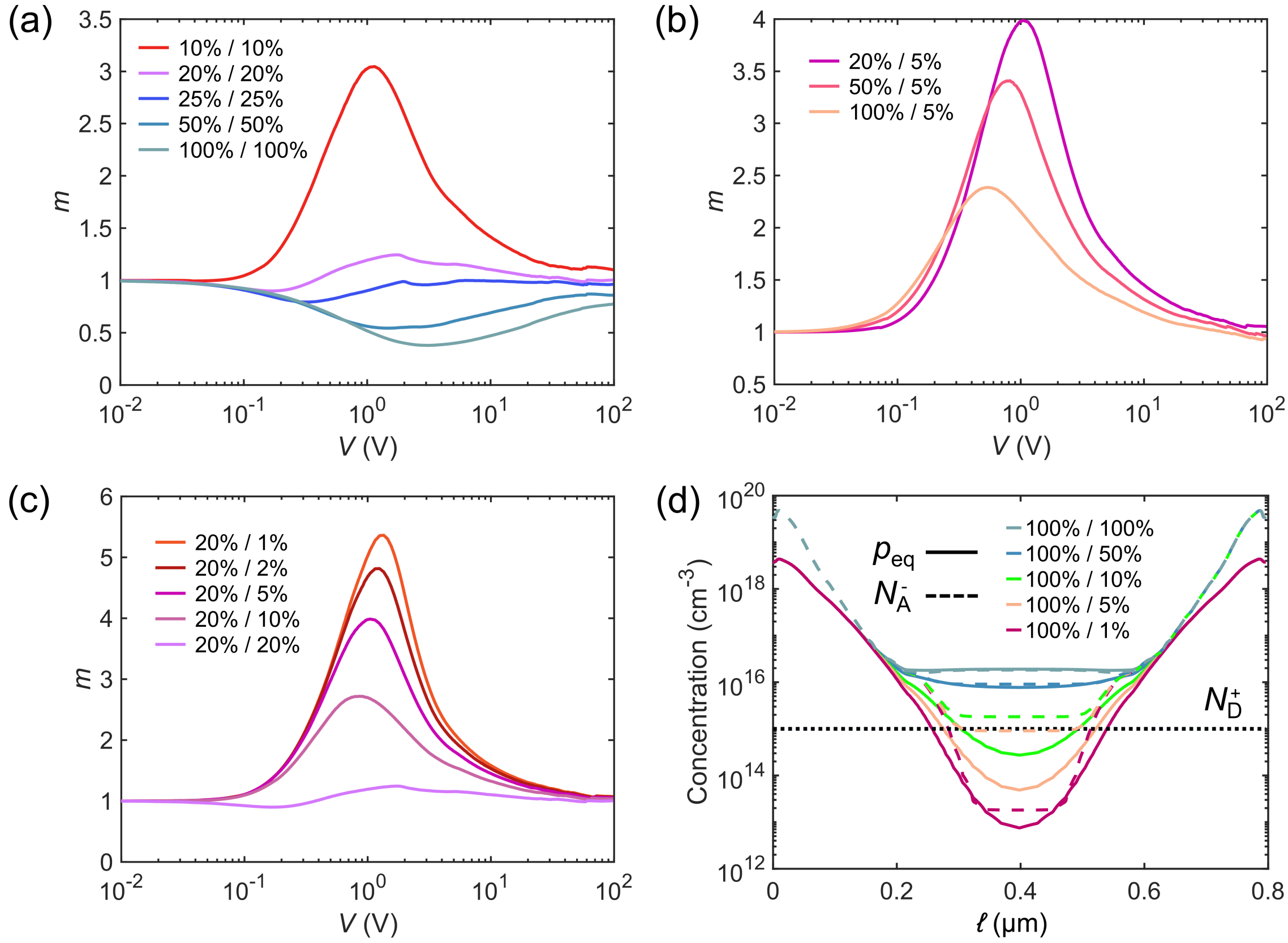

**Figure S20.** TCAD simulations investigating deactivation of implanted dopants. Simulations are performed for a $p$-type TLM structure ($L = 300$ nm, $L_c = 200$ nm, $N_D^+ = 10^{15}$ cm$^{-3}$, $N_{tp} = 10^{12}$ cm$^{-2}$ eV$^{-1}$), assuming different combinations of dopant deactivation under the contacts / in the channel region. Local power-law exponent is plotted (a) assuming the implanted profile is spatially uniform across the device, (b) accounting for varying degrees of dopant deactivation beneath the contacts, and (c) accounting for varying degrees of dopant deactivation in the channel region. (d) The equilibrium hole concentration $p_{eq}$ (solid line), the ionized (implanted) acceptor concentration $N_A^-$ (dashed line) and the ionized (background) donor concentration $N_D^+$ (dotted line) are plotted assuming all dopants are active beneath the contacts, and for varying fractions of active dopants in the plasma-exposed channel region. The concentrations are taken in the center of the device, 130 nm below the etched surface, similarly to the approach taken in Figure 4b.

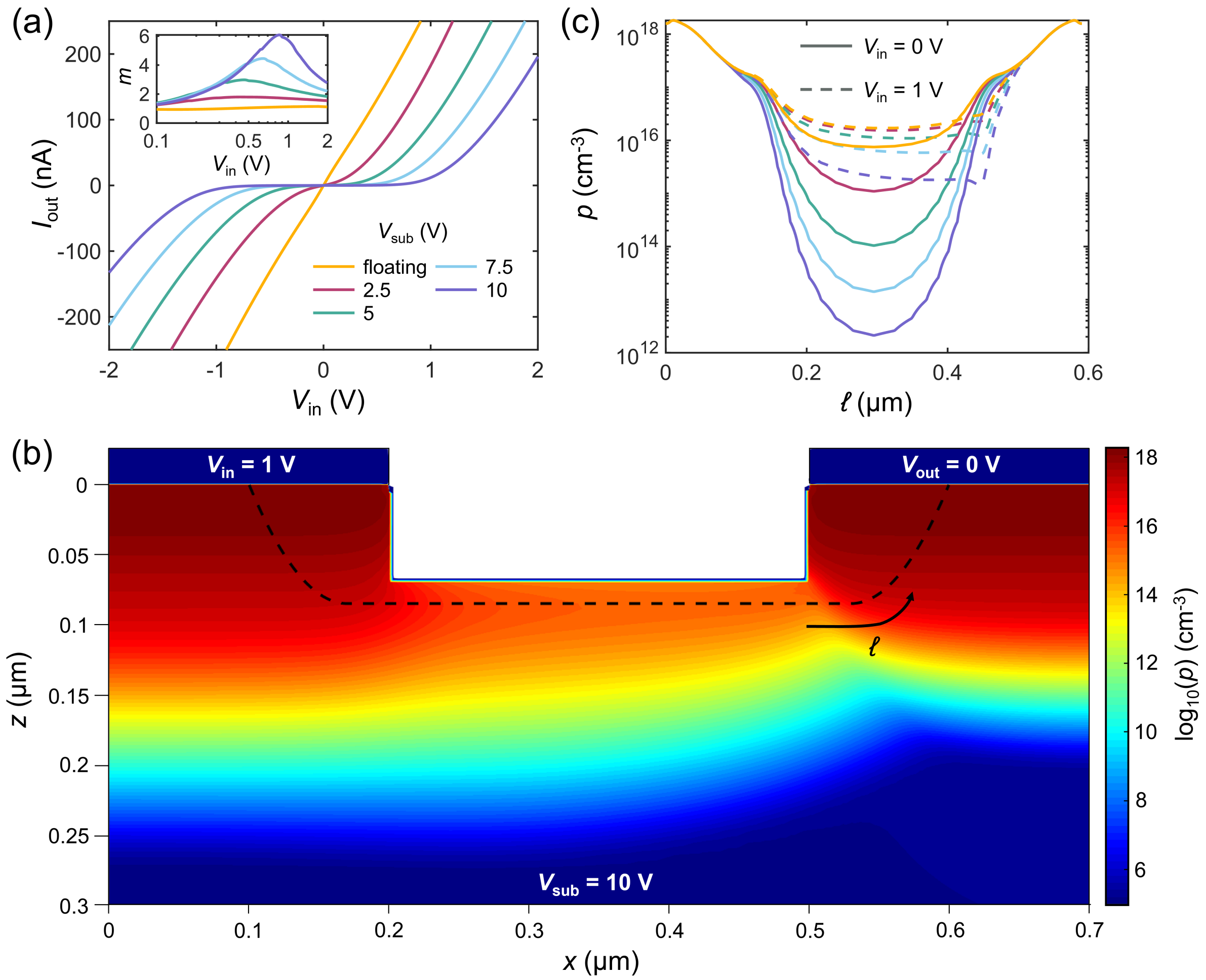

**Figure S21.** TCAD simulations investigating the influence of the substrate potential on the *I*–*V* characteristics of RNPUs. (a) *I*–*V* characteristics for a floating substrate (3 μm thick, not shown here) and increasing substrate voltages $V_{\mathrm{sub}}$ ($L$ = 300 nm, $L_{\mathrm{c}}$ = 200 nm, $N_{\mathrm{D}}^{+} = 10^{15}$ cm$^{-3}$, $N_{\mathrm{tp}} = 10^{12}$ cm$^{-2}$ eV$^{-1}$). Inset: Local power-law exponent $m(V)$ for varying $V_{\mathrm{sub}}$. (b) Free hole concentration $p$ with voltage $V_{\mathrm{in}}$ = 1 V on the sourcing input electrode, $V_{\mathrm{out}}$ = 0 V on the grounded output electrode, and $V_{\mathrm{sub}}$ = 10 V on the substrate. Hole current flows along path $\ell$. (c) Hole concentration $p$ along path $\ell$ in (b) for increasing $V_{\mathrm{sub}}$. Solid and dashed lines indicate $p$ at $V_{\mathrm{in}}$ = 0 V and $V_{\mathrm{in}}$ = 1 V, respectively.

# References

1. Victory Device (Version 1.24.0.R). Silvaco, Inc. Santa Clara, CA, USA. 2024. Available at: https://silvaco.com/.

2. C. R. Helms, and E. H. Poindexter. The Silicon-Silicon Dioxide System: Its Microstructure And Imperfections. *Reports on Progress in Physics*, *57*, no. 8, 791–852, 1994. https://doi.org/10.1088/0034-4885/57/8/002.

3. P. M. Lenahan, T. D. Mishima, J. Jumper, T. N. Fogarty, and R. T. Wilkins. Direct Experimental Evidence For Atomic Scale Structural Changes Involved In The Interface-Trap Transformation Process. *IEEE Transactions on Nuclear Science*, *48*, no. 6, 2131–2135, 2001. https://doi.org/10.1109/23.983184.